\documentclass[twocolumn,twocolappendix]{aastex631}

\usepackage{graphicx}	
\usepackage{amsmath}	
\usepackage{subfigure}
\usepackage{natbib}
\usepackage{color}
\usepackage{tabularx}
\usepackage{longtable}
\usepackage{bm}
\usepackage{threeparttable}
\usepackage{enumerate}
\usepackage[figuresright]{rotating}
\usepackage[normalem]{ulem}
\usepackage{verbatim}
\usepackage[T1]{fontenc}
\usepackage{booktabs}
\usepackage{multirow}

\newcommand{\code}[1]{{\texttt{#1}}}

\newcommand{\gaia}{{\it Gaia}}

\graphicspath{{./}{figures/}}

\begin{document}

\title{Dawn of the Gaia Era: Revisiting Cold Jupiters Orbiting M Dwarfs with Astrometry+Doppler Data Joint Analysis}

\correspondingauthor{Tianjun Gan, Fabo Feng}
\email{tianjungan@gmail.com, ffeng@sjtu.edu.cn}

\author[0009-0000-9341-3442]{Yifan Xuan}
\affil{State Key Laboratory of Dark Matter Physics, Tsung-Dao Lee Institute \& School of Physics and Astronomy, Shanghai Jiao Tong University, Shanghai 201210, China}

\author[0000-0002-4503-9705]{Tianjun~Gan}
\affil{Instituto de Astrof\'{i}sica de Canarias (IAC), E-38205 La Laguna, Tenerife, Spain}
\affil{Departamento de Astrof\'isica, Universidad de La Laguna (ULL), E-38206 La Laguna, Tenerife, Spain}

\author[0000-0001-6039-0555]{Fabo Feng}
\affil{State Key Laboratory of Dark Matter Physics, Tsung-Dao Lee Institute \& School of Physics and Astronomy, Shanghai Jiao Tong University, Shanghai 201210, China}

\author[0000-0002-4355-9838]{In-Gu Shin} 
\affil{Department of Astronomy, Westlake University, Hangzhou 310030, Zhejiang Province, People’s Republic of China}

\author[0000-0001-6753-4611]{Guang-yao Xiao}
\affil{State Key Laboratory of Dark Matter Physics, Tsung-Dao Lee Institute \& School of Physics and Astronomy, Shanghai Jiao Tong University, Shanghai 201210, China}

\author[0009-0008-8931-1513]{Bryant Randolph}
\affil{State Key Laboratory of Dark Matter Physics, Tsung-Dao Lee Institute \& School of Physics and Astronomy, Shanghai Jiao Tong University, Shanghai 201210, China}

\author[0000-0001-8317-2788]{Shude Mao}
\affil{Department of Astronomy, Westlake University, Hangzhou 310030, Zhejiang Province, People’s Republic of China}

\author[0000-0003-0987-1593]{Enric Pall\'{e}} 
\affil{Instituto de Astrof\'{i}sica de Canarias (IAC), E-38205 La Laguna, Tenerife, Spain}
\affil{Departamento de Astrof\'isica, Universidad de La Laguna (ULL), E-38206 La Laguna, Tenerife, Spain}

\author[0000-0002-8388-6040]{Pedro J. Amado}
\affil{Instituto de Astrof\'isica de Andaluc\'ia, CSIC, Glorieta de la Astronom\'ia s/n, 18008 Granada, Spain}

\author[0000-0001-7119-1105]{Alan~P.~Boss}
\affil{Earth and Planets Laboratory, Carnegie Institution for Science, Washington, DC 20015, USA}

\author[0000-0003-1305-3761]{R.~Paul~Butler}
\affil{Earth and Planets Laboratory, Carnegie Institution for Science, Washington, DC 20015, USA}

\author[0000-0002-7349-1387]{Jos\'e~A.~Caballero}
\affil{Centro de Astrobiolog\'ia (CSIC-INTA), Camino bajo del castillo s/n, ESAC Campus, 28692 Villanueva de la Cañada, Madrid, Spain}

\author[0000-0002-5226-787X]{Jeffrey~D.~Crane}
\affil{Observatories of the Carnegie Institution for Science, 813 Santa Barbara Street, Pasadena, CA 91101, USA}

\author[0000-0002-2482-0180]{Zahra Essack}
\affil{Department of Physics and Astronomy, The University of New Mexico, 210 Yale Blvd NE, Albuquerque, NM 87106, USA}

\author[0000-0003-0433-3665]{Hugh R. A. Jones}
\affiliation{Centre for Astrophysics Research, University of Hertfordshire, College Lane, AL10 9AB, Hatfield, UK}

\author[0000-0003-0061-518X]{Juan Carlos Morales}
\affil{Institut de Ci\`encies de l’Espai (ICE, CSIC), Campus UAB, c/ Can Magrans s/n, E-08193 Bellaterra, Spain}
\affil{Institut d'Estudis Espacials de Catalunya (IEEC), Edifici RDIT, Campus PMT-UPC, 08860 Castelldefels (Barcelona), Spain}

\author[0000-0002-3302-1962]{Andreas Quirrenbach}
\affil{Landessternwarte, Zentrum f\"ur Astronomie der Universit\"at Heidelberg, K\"onigstuhl 12, 69117 Heidelberg, Germany}

\author[0000-0002-0460-8289]{Sabine Reffert}
\affil{Landessternwarte, Zentrum f\"ur Astronomie der Universit\"at Heidelberg, K\"onigstuhl 12, 69117 Heidelberg, Germany}

\author[0000-0003-1242-5922]{Ansgar Reiners}
\affil{Institut f\"ur Astrophysik und Geophysik, Georg-August-Universit\"at G\"ottingen, Friedrich-Hund-Platz 1, 37077 G\"ottingen, Germany}

\author[0000-0002-6689-0312]{Ignasi Ribas}
\affil{Institut de Ci\`encies de l’Espai (ICE, CSIC), Campus UAB, c/ Can Magrans s/n, E-08193 Bellaterra, Spain}
\affil{Institut d'Estudis Espacials de Catalunya (IEEC), Edifici RDIT, Campus PMT-UPC, 08860 Castelldefels (Barcelona), Spain}

\author[0000-0002-7670-670X]{Malena Rice}
\affil{Department of Astronomy, Yale University, New Haven, CT 06511, USA}

\author[0000-0002-8681-6136]{Stephen A. Shectman}
\affil{Observatories of the Carnegie Institution for Science, 813 Santa Barbara Street, Pasadena, CA 91101, USA}

\author[0009-0008-2801-5040]{Johanna~K.~Teske}
\affil{Earth and Planets Laboratory, Carnegie Institution for Science, Washington, DC 20015, USA}

\author[0000-0002-6937-9034]{Sharon X. Wang}
\affil{Department of Astronomy and Tsinghua Centre for Astrophysics, Tsinghua University, Beijing 100084, People's Republic of China}

\author[0000-0002-6532-4378]{Mathias Zechmeister}
\affil{Institut f\"ur Astrophysik und Geophysik, Georg-August-Universit\"at G\"ottingen, Friedrich-Hund-Platz 1, 37077 G\"ottingen, Germany}



\begin{abstract}

We perform a joint analysis of ground-based radial velocities (RVs), Hipparcos Intermediate Astrometric Data, and two-epoch Gaia astrometry for known M dwarfs hosting cold Jupiters discovered through long-term spectroscopic monitoring. 
Among a sample of 25 planets in 19 systems, we derive robust absolute masses and orbital solutions for 13 planets in 11 systems with companion inclination uncertainties lower than 30 degrees. These include first mass measurements for five planets: GJ 328b, GJ 649b, GJ 849b and c, and GJ 9066c. We also reclassify one object, HIP 79431B, as a brown dwarf. We further compare the mass ratios of these 13 cold Jupiter systems with those of giant planets orbiting M stars characterized via transit, microlensing, and direct imaging.
Without correcting for detection bias, we find that (1) there is likely a deficiency of high-mass-ratio hot Jupiters around M dwarfs compared with cold Jupiters, and (2) the local cold Jupiters orbiting M stars found by RV observations exhibit a similar mass ratio distribution as the counterparts in the Galactic bulge found by microlensing. Cold Jupiters orbiting stars with masses above $0.65~M_\odot$ tend to be more massive than those around M dwarfs. Our study establishes a sample of local cold Jupiters with precise mass determinations around M dwarfs, and reveal a tip of the iceberg in the upcoming Gaia DR4 and DR5.

\end{abstract}



\section{Introduction} \label{sec:intro}

About two thousand giant planets have been detected to date \citep{Akeson2013,Christiansen2025}, yet only a small fraction of them orbit M dwarfs, even though these low-mass stars are the most abundant stellar population in the Milky Way \citep{Henry2018,Golovin2023}. With mass ratios approaching those of brown dwarfs orbiting Sun-like stars and even some binary star systems, the formation mechanism of giant planets around M dwarfs remains under debate.

One of the leading planet formation theories, core accretion, postulates that the process begins with a solid core, which can then grow into a giant planet by rapidly accreting gas \citep{Pollack1996,Ida2004II,Mordasini2008}. Several lines of observational evidence suggest that giant planets around M dwarfs likely form through this scenario. For example, \cite{Gan2023} measured an occurrence rate of 
$0.27\pm0.09\%$ for hot Jupiters orbiting early-type M dwarfs \citep[see also][]{Bryant2023,Glusman2026}, roughly three times lower than that around Sun-like stars \citep{Petigura2018,Zhou2019,Beleznay2022}. Within the core accretion framework, the lower surface density of M dwarf protoplanetary disks compared to those around Sun-like stars impedes the formation of a sufficiently massive core to initiate runaway gas accretion before the disk dissipates \citep{Laughlin2004,Ida2005,Kennedy2008}, which could account for the difference in occurrence rate. Moreover, formation of hot Jupiters is favored around metal-rich host stars, though the correlation seems less significant for cold Jupiters \citep{Maldonado2020,Gan2025a,Morales2025}. Stellar metallicity is thought to correlate with the amount of solid material in the protoplanetary disk for planet formation \citep{Santos2004,Fischer2005,Sousa2011}. One would therefore naturally expect more giant planets around metal-rich stars if they formed through this bottom-up channel. Although core accretion is efficient beyond the snow line where dust is abundant \citep{Ida2005}, outer massive companions could excite the orbits of inner giant planets and drive them inward to compact orbits \citep{Naoz2016,Dawson2018}. Recent work from \cite{Gan2026b} found that the stellar multiplicity of M dwarfs hosting hot Jupiters is about six times higher than the fraction of field M dwarfs, indicating that high-eccentricity migration may play an important role in producing short-period transiting hot Jupiters around low-mass stars \citep[e.g.,][]{Gantoi530,Gan2023TOI4201,Jordan2022,Hartman2024,Kanodia2024,Dransfield2026,Frensch2026}.

Nevertheless, a few Jovian-like planets orbiting very-low-mass stars with $M_\ast\lesssim0.2~M_\odot$ remain challenging to explain via the core accretion route \citep[e.g.,][]{Morales2019,Quirrenbach2022,Bryant2025}. An alternative channel, gravitational instability \citep{Boss1997,Boss2000}, is speculated to be responsible for forming such extreme-mass-ratio systems. In this scenario, a gravitationally unstable disk can condense and fragment into massive giant planets at a few AUs, a semi-major axis range to which the microlensing and RV methods are sensitive. While the majority of cold Jupiters orbiting M dwarfs have so far been discovered by microlensing \citep{Mao1991,Gould1992}, degeneracies in orbital solutions are common, such as close/wide degeneracy \citep{Griest1998}, which causes considerable uncertainties in host-planet separations. Moreover, the microlensing technique provides projected separations of planetary systems instead of true semi-major axes in most cases \citep[e.g.,][]{Shin2023,Shin2024,Shin2026}. As for the RV technique, a handful of wide-orbit large planets orbiting M dwarfs have also been found through blind surveys \citep[e.g.,][]{Butler2006catalog,Howard2010,Quirrenbach2022}. However, long-term monitoring is required to cover adequate orbital phases, making planet searches and identifications inefficient. Additionally, the $\sin i$ degeneracy leaves the absolute mass of the RV-detected planets unknown, hence there is always a risk that some ``planets'' are sub-stellar or stellar companions \citep[e.g.,][]{Kiefer2019,Xiao2023}. 

As a complement to the RV technique that traces the line-of-sight motion, the astrometric method measures the wobble of stars on the sky plane \citep{Curiel2022}. The six-dimensional kinematic information of astrometry allows full-orbit determination for the companion, enabling us to disentangle the orbital inclination and uniquely specify its true mass. With the upcoming Gaia DR4, thousands of wide-orbit giant planets orbiting low-mass M dwarfs are anticipated to be discovered \citep{Sozzetti2014,Perryman2014,Lammers2026}, providing a golden sample for comparison with local transiting hot Jupiters as well as microlensed cold Jupiters in the Galactic Bulge to infer their formation history. Though the timeseries astrometry of Gaia DR4 is not available yet, joint analyses of RV and absolute astrometry data have already been widely applied in several works \citep[e.g.,][]{Brandt2019,Li2021,Feng2022,Xiao2023,An2025,Piccinini2026,Ren2026}. 

Before stepping into the Gaia DR4 era, here we revisited all RV-detected giant planets around M dwarfs. We perform a homogeneous joint analysis of ground-based long-term RVs together with both Hipparcos and Gaia absolute astrometry \citep{Perryman1997,vanLeeuwen2007,Gaia2018_DR2summary,Gaia2023_DR3summary}. We determine the absolute planet masses of 13 cold Jupiters orbiting 11 M stars, and five of them (GJ 328b, GJ 649b, GJ 849b and c, and GJ 9066c) are the first measurements. In addition to planetary systems, we also confirm that the RV-detected cold Jupiter around HIP 79431 \citep{Apps2010} is a brown dwarf companion. The rest of the paper is structured as follows. We start with sample construction in Section~\ref{sample_construction} and summarize the astrometry and RV data we used in Section~\ref{data}. Section~\ref{jointfit} describes the details of the RV+Astrometry joint fit analysis, followed by the results for each system in Section~\ref{results}. We discuss our findings in Section~\ref{discussion} before we conclude in Section~\ref{conclusion}.

\section{Sample}\label{sample_construction}

To construct a sample of wide-orbit giant planets orbiting M dwarfs for RV+Astrometry analysis, we retrieved a list of all confirmed planets from the NASA exoplanet archive \citep{Akeson2013, Christiansen2025} UT 2026 October 1. We then filtered out RV-detected giant planets orbiting M dwarfs through the following selection criteria:
\begin{enumerate}[(1)]

\item Stellar mass: $M_\ast \leq 0.65~M_\odot$;
\item Stellar effective temperature: $T_{\rm eff}\leq 4000~K$;
\item Minimum planet mass: $0.2\leq M_{p}\sin i\leq13.6~M_J$;
\item Discovery Method: Radial Velocity.

\end{enumerate}

In addition to previously published systems, we include a new discovery, HIP~88962b, a massive giant planet orbiting an M0V star with a period of about 19 years, which is discussed in detail in a separate paper \citep{Boss2026}. Eventually, our sample consists of 25 cold Jupiters in 19 systems, all of which have orbital periods longer than 30 days. 

Next, we cross-matched the planet catalog with the cool dwarf list \citep[cooldwarfs\_v8;][]{Muirhead2018} embedded in the TESS input Catalog v8.2 \citep[TIC;][]{Stassun2019ticv8} to obtain a homogeneous set of stellar parameters. 
In brief, the authors computed the absolute $K_S$ magnitude from the {\it 2MASS} catalog \citep{Cutri:2003,skrutskie2006}, and utilized the mass-luminosity relation to estimate stellar mass \citep{Delfosse2000,Benedict2016,Mann2019}. 
Similarly, stellar radius was determined using the empirical polynomial relation reported by \cite{Mann2015}. Broadband optical photometry $G_{\rm BP}$ and $G_{\rm RP}$ from \gaia\ \citep{Gaia2018_DR2summary} was adopted to compute the stellar effective temperature $T_{\rm eff}$ using custom relations based on photometrically calibrated spectra described in \cite{Mann2013b}. 
However, both stellar mass and radius of GJ 508.2 reported in TIC have large uncertainties, and $T_{\rm eff}$ is not provided, likely due to the large uncertainty in its $K_S$ magnitude. We therefore adopt the stellar parameters from \citet{Morales2025}. Moreover, the effective temperature of GJ 832 is taken from \cite{Wittenmyer2014}, due also to the lack of result in TIC. Table~\ref{stellar_parameters} lists the stellar parameters of these 19 systems.

\begin{sidewaystable*}[p]
\centering
\raisebox{-5.0cm}[0pt][0pt]{%
\hspace*{-1.0cm}%
\begin{minipage}{0.98\textheight}

    \caption{Stellar parameters of 19 M dwarfs hosting cold Jupiters}
    \label{stellar_parameters}
    
    \setlength{\tabcolsep}{6.0pt}
    \renewcommand{\arraystretch}{2.5}

    \resizebox{\linewidth}{!}{%
    \fontsize{12pt}{13pt}\selectfont
    \begin{tabular}{@{}cccccccccccccc@{}}
        \hline\hline
         Target &HIP ID &$\alpha_{\rm J2016.0}$ &$\delta_{\rm J2016.0}$ &$\varpi$ &$\mu_{\alpha}$  &$\mu_{\delta}$ &RV &$G$ &$G_{\rm BP}$ &$G_{\rm RP}$ &$M_{\ast}$ &$R_{\ast}$ &$T_{\rm eff}$ \\
        & & & &$[{\rm mas}]$ &$[{\rm mas\ yr^{-1}}]$ &$[{\rm mas\ yr^{-1}}]$ &$[{\rm km \ s^{-1}}]$ &[mag] &[mag] &[mag] &$[{\rm M_{\odot}}]$ &$[{\rm R_{\odot}}]$ &$[{\rm K}]$ \\\hline
        GJ 179 &22627 &$04^{\rm h}52^{\rm m}05.90^{\rm s}$ &$+06^\circ 28' 30.69''$ &$80.56\pm0.02$ &$153.62\pm0.03$ &$-306.05\pm0.02$ &$-8.28\pm0.26$ &$10.7140\pm0.0004$ &$12.2307\pm0.0011$ &$9.5074\pm0.0007$ &$0.364\pm0.02$ &$0.376\pm0.011$ &$3315\pm157$ \\
        GJ 317 & -- & $08^{\rm h}40^{\rm m}58.67^{\rm s}$ & $-23^\circ 27' 09.71''$ & $65.88\pm0.04$ & $-461.34\pm0.04$ & $805.22\pm0.04$ & $87.11\pm0.18$ & $10.7651\pm0.0007$ & $12.2327\pm0.0018$ & $9.5689\pm0.0009$ & $0.421\pm0.02$ & $0.427\pm0.013$ & $3347\pm157$ \\
        GJ 328 & 43790 & $08^{\rm h}55^{\rm m}07.67^{\rm s}$ & $+01^\circ 32' 30.68''$ & $48.74\pm0.02$ & $44.94\pm0.02$ & $-1045.88\pm0.01$ & $-4.27\pm0.13$ & $9.2895\pm0.0002$ & $10.1863\pm0.0006$ & $8.3539\pm0.0003$ & $0.628\pm0.02$ & $0.651\pm0.019$ & $3923\pm157$  \\
        GJ 463 & 60398 & $12^{\rm h}22^{\rm m}58.52^{\rm s}$ & $+64^\circ 01' 56.94''$ & $54.45\pm0.02$ & $-673.12\pm0.02$ & $373.31\pm0.02$ & $8.92\pm0.16$ & $10.5524\pm0.0003$ & $11.8252\pm0.0010$ & $9.4291\pm0.0004$ & $0.472\pm0.02$ & $0.475\pm0.014$ & $3487\pm157$ \\
        GJ 508.2 & 65143 & $13^{\rm h}20^{\rm m}58.67^{\rm s}$ & $+34^\circ 16' 39.44''$ & $60.03\pm0.02$ & $483.88\pm0.01$ & $-297.07\pm0.02$ & $-35.80\pm0.14$ & $9.7727\pm0.0003$ & $10.8363\pm0.0008$ & $8.7456\pm0.0004$ & $0.54\pm0.12$ & $0.53\pm0.12$ & $3720\pm12$  \\
        GJ 649 & 83043 & $16^{\rm h}58^{\rm m}08.71^{\rm s}$ & $+25^\circ 44' 30.84''$ & $96.23\pm0.02$ & $-115.31\pm0.02$ & $-508.09\pm0.03$ & $3.76\pm0.13$ & $8.8236\pm0.0004$ & $9.9003\pm0.0011$ & $7.7879\pm0.0006$ & $0.514\pm0.02$ & $0.517\pm0.015$ & $3674\pm157$  \\
        GJ 676A & 85647 & $17^{\rm h}30^{\rm m}10.76^{\rm s}$ & $-51^\circ 38' 16.09''$ & $62.58\pm0.03$ & $-258.76\pm0.03$ & $-185.12\pm0.02$ & $-39.82\pm0.14$ & $8.8669\pm0.0003$ & $9.8094\pm0.0008$ & $7.9065\pm0.0007$ & $0.626\pm0.02$ & $0.649\pm0.02$ & $3851\pm157$ \\
        GJ 832 & 106440 & $21^{\rm h}33^{\rm m}33.90^{\rm s}$ & $-49^\circ 00' 45.47''$ & $201.33\pm0.02$ & $-45.92\pm0.02$ & $-816.87\pm0.02$ & $12.72\pm0.13$ & $7.7399\pm0.0003$ & $8.9025\pm0.0007$ & $6.6626\pm0.0005$ & $0.437\pm0.02$ & $0.442\pm0.013$ & $3472$ \\
        GJ 849 & 109388 & $22^{\rm h}09^{\rm m}41.56^{\rm s}$ & $-04^\circ 38' 27.01''$ & $113.44\pm0.03$ & $1132.58\pm0.04$ & $-22.16\pm0.04$ & $-15.61\pm0.16$ & $9.2372\pm0.0007$ & $10.6117\pm0.0023$ & $8.0700\pm0.0009$ & $0.461\pm0.02$ & $0.464\pm0.014$ & $3407\pm157$  \\
        GJ 876 & 113020 & $22^{\rm h}53^{\rm m}17.79^{\rm s}$ & $-14^\circ 16' 00.08''$ & $214.04\pm0.04$ & $957.71\pm0.04$ & $-673.60\pm0.03$ & $-2.09\pm0.15$ & $8.8751\pm0.0008$ & $10.4530\pm0.0024$ & $7.6437\pm0.0011$ & $0.337\pm0.02$ & $0.352\pm0.011$ & $3271\pm157$ \\
        GJ 1148 & 57050 & $11^{\rm h}41^{\rm m}43.80^{\rm s}$ & $+42^\circ 45' 05.67''$ & $90.69\pm0.03$ & $-575.53\pm0.03$ & $-89.80\pm0.02$ & $-8.97\pm0.22$ & $10.5821\pm0.0005$ & $12.1394\pm0.0021$ & $9.3610\pm0.0010$ & $0.344\pm0.02$ & $0.359\pm0.011$ & $3287\pm157$ \\
        GJ 3512 & -- & $08^{\rm h}41^{\rm m}19.58^{\rm s}$ & $+59^\circ 29' 29.97''$ & $105.29\pm0.03$ & $-260.28\pm0.02$ & $-1279.56\pm0.03$ & $7.13\pm2.45$ & $13.0823\pm0.0008$ & $15.3833\pm0.0036$ & $11.7121\pm0.0017$ & $0.128\pm0.02$ & $0.156\pm0.005$ & $2902\pm157$ \\
        GJ 9066 & -- & $02^{\rm h}00^{\rm m}14.16^{\rm s}$ & $+13^\circ 02' 38.66''$ & $223.73\pm0.07$ & $1096.46\pm0.07$ & $-1771.53\pm0.06$ & $-28.29\pm0.25$ & $10.6811\pm0.0008$ & $12.5719\pm0.0025$ & $9.3837\pm0.0013$ & $0.15\pm0.02$ & $0.18\pm0.005$ & $3102\pm157$ \\
        GJ 9773 & 109638 & $22^{\rm h}12^{\rm m}36.06^{\rm s}$ & $+08^\circ 33' 00.87''$ & $62.88\pm0.02$ & $106.38\pm0.03$ & $-671.15\pm0.04$ & $8.19\pm0.17$ & $10.9082\pm0.0005$ & $12.2297\pm0.0017$ & $9.7698\pm0.0005$ & $0.367\pm0.02$ & $0.378\pm0.012$ & $3453\pm157$  \\
        HIP 12961 & 12961 & $02^{\rm h}46^{\rm m}43.23^{\rm s}$ & $-23^\circ 05' 09.55''$ & $177.93\pm0.04$ & $294.72\pm0.01$ & $140.97\pm0.01$ & $32.49\pm0.14$ & $9.6029\pm0.0003$ & $10.4729\pm0.0006$ & $8.6867\pm0.0003$ & $0.62\pm0.083$ & $0.661\pm0.069$ & $3986\pm125.8$ \\
        HIP 79431 & 79431 & $16^{\rm h}12^{\rm m}41.82^{\rm s}$ & $-18^\circ 52' 35.24''$ & $75.88\pm0.06$ & $36.45\pm0.03$ & $-214.02\pm0.02$ & $-4.93\pm0.19$ & $10.2506\pm0.0005$ & $11.6122\pm0.0015$ & $9.0945\pm0.0008$ & $0.477\pm0.02$ & $0.479\pm0.014$ & $3420\pm157$ \\
        HIP 88962 & 88962 & $18^{\rm h}09^{\rm m}33.34^{\rm s}$ & $-12^\circ 02' 23.29''$ & $36.67\pm0.01$ & $67.01\pm0.02$ & $-206.82\pm0.01$ & $-1.31\pm0.31$ & $9.8511\pm0.0005$ & $10.7129\pm0.0011$ & $8.9426\pm0.0009$ & $0.62\pm0.084$ & $0.683\pm0.068$ & $3993\pm122.1$ \\
        TOI-756 & -- & $12^{\rm h}48^{\rm m}24.89^{\rm s}$ & $-45^\circ 28' 13.68''$ & $11.61\pm0.02$ & $-216.50\pm0.02$ & $29.20\pm0.01$ & $15.36\pm1.20$ & $13.6773\pm0.0004$ & $14.8995\pm0.0019$ & $12.5804\pm0.0009$ & $0.506\pm0.02$ & $0.508\pm0.015$ & $3533\pm157$ \\
        TYC 2187-512-1 & -- & $21^{\rm h}22^{\rm m}06.41^{\rm s}$ & $+22^\circ 55' 55.03''$ & $64.58\pm0.02$ & $108.77\pm0.02$ & $117.79\pm0.02$ & $5.45\pm0.20$ & $9.7752\pm0.0004$ & $10.8307\pm0.0015$ & $8.7556\pm0.0006$ & $0.498\pm0.02$ & $0.499\pm0.015$ & $3705\pm157$ \\
        \hline
    \end{tabular}
    }
\par
\vspace{3mm}
\noindent
\hspace*{65pt}%
\parbox{\dimexpr\linewidth-65pt\relax}{%
\footnotesize
\raggedright
\textbf{Notes---}
[1] The right ascension, declination, parallax, proper motions, absolute radial velocity (RV), and Gaia $G$, $G_{\rm BP}$, and $G_{\rm RP}$ magnitudes of the M dwarfs in our sample are adopted from Gaia DR3 \citep{Gaia2023_DR3summary};
}
\end{minipage}
\par
}%
\end{sidewaystable*}

\section{Data} \label{data}

\subsection{Absolute Astrometric Data}
In this study, we use astrometric data from both the Hipparcos and Gaia catalogs, including positions in right ascension ($\alpha$) and declination ($\delta$), proper motions ($\mu_{\alpha}$ and $\mu_{\delta}$), and parallax ($\varpi$).
Specifically, we incorporate Hipparcos catalogue data and intermediate astrometric data (IAD; also referred to as abscissa data) from the DVD version of \cite{vanLeeuwen2007}, and the Gaia second and third data releases (GDR2 and GDR3; \citealt{Gaia2018_DR2summary,Gaia2023_DR3summary}) when available.
Since Gaia IAD is not publicly available until December of 2026, we use the Gaia Observation Schedule Tool\footnote{\url{https://gaia.esac.esa.int/gost/index.jsp}}(GOST) to obtain the Gaia observing cadence and scan-angle geometry, which are combined with our astrometric model to simulate along-scan measurements.
We have taken into account the publicly available Gaia observational gaps\footnote{\url{https://www.cosmos.esa.int/web/gaia/dr3-data-gaps}
} (e.g., dead times and rejected observations) when modeling Gaia IAD.

We note that the GDR2 and GDR3 astrometry are not strictly independent because their underlying observations partially overlap. 
Since the cross-release covariance is unavailable, we treat the two catalog solutions as independent in our likelihood. 
\cite{Zhao2026} empirically tested the impact of this approximation using systems with Gaia non-single-star orbital solutions and found no significant systematic bias in the inferred orbital inclinations. 
We calibrated GDR2 astrometry to the GDR3 reference frame using the recommended parameters derived in \cite{Feng2024}\footnote{A python script is publicly available at \url{https://github.com/ruiyicheng/Download_HIP_Gaia_GOST}}. 
This correction mitigates spurious astrometric offsets between the two Gaia catalogs and prevents overestimation of the orbital astrometric signal \citep{Feng2024,Thompson2026}. For stars within the magnitude–color calibration grid, we applied the coefficients corresponding to their Gaia $G$ magnitude and $BP-RP$ color. Seven red stars fall outside the grid at their respective magnitudes: GJ 179, GJ 317, GJ 463, GJ 1148, GJ 9066, GJ 9773, and HIP 79431. 
For these stars, we instead adopted the global calibration coefficients listed in Table 1 of \cite{Feng2024}: the NST coefficients for six stars with $G\geq10.5$ and the SS coefficients for HIP 79431 ($G<10.5$). 

For five systems that do not have Hipparcos IAD including GJ 317, GJ 3512, GJ 9066, TOI-756 and TYC 2187-512-1, we carry out the analysis with GDR2+GDR3 data. The astrometric anomaly between Hipparcos and Gaia, corresponding to a temporal baseline of $\sim25$ yr, is primarily sensitive to long-period companions \citep{Brandt2018, Brandt2021HGCA, Kervella2022, Feng2024}, whereas the shorter-term anomaly between GDR2 and GDR3 is more sensitive to companions on orbits with periods of order $\sim10^3$ days \citep{Xiao2026ApJ}. 


\subsection{Radial Velocities}

We made use of archival RV data from several instruments including CARMENES installed at the 3.5~m telescope at the Observatorio de Calar Alto in Almería \citep{Quirrenbach2014}, the High Accuracy Radial velocity Planet Searcher (HARPS) as well as the Near-InfraRed Planet Searcher (NIRPS) spectrograph on the ESO La Silla 3.6~m telescope \citep{Mayor2003,Bouchy2025} and its twin instrument HARPS-N in the Northern Hemisphere installed at the 3.6~m Italian Telescopio Nazionale Galileo telescope \citep{Cosentino2012}, the high-resolution echelle spectrometer (HIRES) on the 10~m Keck I telescope \citep{Vogt1994, Butler2017, Teklu2025}, the High Resolution Spectrograph (HRS) mounted on the 10~m Hobby-Eberly Telescope \citep{Tull1998}, the InfraRed Doppler spectrograph (IRD) installed on the the 8.2~m Subaru telescope \citep{Kotani2018}, the Carnegie Planet Finder Spectrograph (PFS) on the 6.5~m Magellan Clay Telescope \citep{Crane2010}, the SOPHIE spectrograph installed on the 1.93~m reflector telescope at the Haute-Provence Observatory \citep{Perruchot2008}, the Tull spectrograph on the 2.72~m Harlan J. Smith telescope of McDonald Observatory \citep{Tull1995} as well as University College London Echelle Spectrograph (UCLES) mounted on the 3.9~m Anglo-Australian Telescope \citep{Diego1990}. 

The CARMENES observations are performed simultaneously in the visible (VIS) and near-infrared (NIR) channels. Here we only use the VIS-band data, denoted as CAR-VIS, which have better precision. Since HIRES and HARPS underwent upgrades in August 2004 \citep{Butler2006catalog} and June 2015 \citep{Lo2015}, resulting in a zero-point velocity offset and improved RV precision, we thus treat the data before and after these upgrades as coming from separate instruments. We denoted the datasets as HIRES-pre, HIRES-post, HARPS-pre and HARPS-post. 

In addition to publicly available data, we also report and incorporate new PFS RVs for GJ 179, GJ 676A, GJ 832, GJ 849, GJ 876 and HIP 12961 and CARMENES RVs for GJ 317, GJ 463 and GJ 508.2. All CARMENES data were reduced with the SpEctrum Radial Velocity AnaLyser \citep[SERVAL;][]{Zechmeister2018}, while the PFS RVs were extracted through the pipeline developed by \cite{Butler1996}. The HARPS RVs are adopted either from the header measured with the standard data reduction pipeline or from the HARPS RVBank archive \citep{Trifonov2020,HARPSRVBank2}. Other data were retrieved from the literature. We summarize the RV information of each target in Table~\ref{Tab:RV}. All RVs used in this work are listed in Table~\ref{RVvalues} and available in its entirety in machine-readable form. 

\begin{table*}
\centering
\caption{RV Data Summary for Our Sample}\label{Tab:RV}
\resizebox{\textwidth}{!}{
\begin{tabular}{llccccllcccc}
\hline \hline
Name & Instrument & $N_{\rm obs} $ & $\langle \sigma_{\rm RV}\rangle$ & Time span & Refs & Name & Instrument & $N_{\rm obs} $ & $\langle \sigma_{\rm RV}\rangle$ & Time span & Refs\\
 & & & (${\rm m\ s}^{-1}$) & (days) & & & & & (${\rm m\ s}^{-1}$) & (days) & \\
\hline
GJ 179 & CAR-VIS & 10 & 1.35 & 1458 & 1 & GJ 849 & CAR-VIS & 60 & 1.53 & 1250 & 1\\
 & HARPS-pre & 22 & 2.73 & 1548 & 2 & & HARPS-N & 94 & 0.81 & 944 & 9\\
 & HARPS-post & 4 & 6.49 & 89 & 2 & & HARPS-pre & 48 & 0.80 & 2824 & 6\\
 & HIRES-pre & 11 & 4.59 & 1436 & 3 & & HARPS-post & 47 & 2.17 & 2810 & 2\\
 & HIRES-post & 67 & 2.22 & 5550 & 3 & & HIRES-pre & 24 & 3.19 & 2591 & 7\\
 & HRS & 14 & 8.35 & 1909 & 3 & & HIRES-post & 111 & 1.52 & 5423 & 7\\
 & PFS & 21 & 2.63 & 1982 & 4 & & PFS & 7 & 0.88 & 3010 & 4\\
GJ 317 & CAR-VIS & 47 & 1.90 & 3001 & 5, 4 & GJ 876 & CAR-VIS & 70 & 1.45 & 1262 & 15, 1\\
 & HARPS-pre & 84 & 1.65 & 1440 & 6 & & HIRES-pre & 140 & 2.11 & 2595 & 7\\
 & HARPS-post & 70 & 4.91 & 2746 & 2 & & HIRES-post & 213 & 2.26 & 5189 & 7\\
 & HIRES-pre & 5 & 4.73 & 1438 & 7 & & HARPS-pre & 257 & 0.71 & 4016 & 6\\
 & HIRES-post & 52 & 2.14 & 5484 & 7 & & HARPS-post & 13 & 0.72 & 1133 & 6\\
 & PFS & 32 & 2.76 & 2520 & 8 & & PFS & 17 & 1.08 & 2108 & 4\\
GJ 328 & HARPS-N & 130 & 0.95 & 1132 & 9 & GJ 1148 & CAR-VIS & 76 & 1.48 & 1194 & 1\\
 & HARPS-post & 11 & 1.09 & 107 & 2 & & HIRES-pre & 8 & 3.90 & 1496 & 7\\
 & HIRES-post & 4 & 3.65 & 1093 & 10 & & HIRES-post & 116 & 2.04 & 5093 & 7\\
 & HRS & 58 & 6.39 & 3753 & 10 & GJ 3512 & CAR-VIS & 227 & 2.82 & 2869 & 5\\
 & SOPHIE & 36 & 4.05 & 1827 & 11 & & IRD & 39 & 3.87 & 1117 & 5\\
 & Tull & 14 & 5.61 & 346 & 10 & GJ 9066 & CAR-VIS & 93 & 1.70 & 1398 & 16\\
GJ 463 & CAR-VIS & 165 & 3.68 & 3459 & 5, 4 & & HIRES-pre & 8 & 3.34 & 1579 & 7\\
 & HIRES-post & 17 & 4.14 & 4377 & 12 & & HIRES-post & 40 & 2.41 & 5537 & 7\\
 & HRS & 53 & 7.19 & 1875 & 12 & GJ 9773 & CAR-VIS & 139 & 1.63 & 3050 & 5\\
GJ 508.2 & CAR-VIS & 32 & 1.52 & 3630 & 5, 4 & HIP 12961 & HARPS-pre & 48 & 1.36 & 2226 & 6\\
GJ 649 & CAR-VIS & 54 & 1.61 & 399 & 1 & & PFS & 50 & 1.42 & 3723 & 4\\
 & HARPS-N & 147 & 0.76 & 917 & 9 & HIP 79431 & HARPS-post & 30 & 3.20 & 1878 & 2\\
 & HARPS-post & 5 & 0.70 & 112 & 6 & & HIRES-post & 13 & 3.09 & 179 & 17\\
 & HIRES-pre & 24 & 1.64 & 1786 & 7 & HIP 88962 & PFS & 48 & 1.73 & 5957 & 18\\
 & HIRES-post & 57 & 1.09 & 4961 & 7 & TOI-756 & HARPS-post & 44 & 5.64 & 421 & 19\\
GJ 676A & HARPS-pre & 128 & 1.14 & 3231 & 6 & & NIRPS & 57 & 16.19 & 477 & 19\\
 & HARPS-post & 9 & 1.06 & 1862 & 6 & TYC 2187-512-1 & CAR-VIS & 94 & 1.55 & 1293 & 16\\
 & PFS & 37 & 1.71 & 4279 & 13, 4 & & & & & & \\
GJ 832 & HARPS-pre & 61 & 0.47 & 3134 & 6 & & & & & & \\
 & HARPS-post & 165 & 1.06 & 2565 & 2 & & & & & & \\
 & PFS & 52 & 0.81 & 5200 & 4 & & & & & & \\
 & UCLES & 39 & 2.59 & 5465 & 14 & & & & & & \\
\hline
\end{tabular}
}
\par 
\vspace{3mm}
\noindent
\hspace*{65pt}%
\parbox{\dimexpr\linewidth-65pt\relax}{%
\footnotesize
\raggedright
\textbf{References—}
(1) \cite{Ribas2023};
(2) ESO Archive;
(3) \cite{Howard2010};
(4) This work;
(5) \cite{Morales2025};
(6) \cite{Trifonov2020};
(7) \cite{Rosenthal2021};
(8) \cite{Feng2020};
(9) \cite{Pinamonti2023};
(10) \cite{Robertson2013};
(11) SOPHIE archive;
(12) \cite{Endl2022};
(13) \cite{Feng2022};
(14) \cite{Wittenmyer2014};
(15) \cite{Trifonov2018};
(16) \cite{Quirrenbach2022}; 
(17) \cite{Apps2010}; 
(18) \cite{Boss2026}; 
(19) \cite{Parc2025}.
\par
}%
\end{table*}

\begin{table*}[htb]
    \centering
    \caption{All radial velocities of 19 stars in our sample}
    \begin{tabular}{ccccc}
        \hline\hline
        BJD       &RV (m~s$^{-1}$) &$\sigma_{\rm RV}$ (m~s$^{-1}$) &Instrument &Object      \\\hline
        2457623.65553 &-7.940 &1.391 &CAR-VIS &GJ 179\\
        2457631.66523 &-0.998 &1.495 &CAR-VIS &GJ 179\\
        2457654.60721 &-6.745 &1.348 &CAR-VIS &GJ 179\\
        2457665.58616 &-1.466 &1.085 &CAR-VIS &GJ 179\\
        2457673.58261 &-5.018 &1.254 &CAR-VIS &GJ 179\\
        2457691.60739 &-5.285 &1.207 &CAR-VIS &GJ 179\\
        2457695.59637 &-4.372 &1.388 &CAR-VIS &GJ 179\\
        2457705.56901 &-1.128 &1.610 &CAR-VIS &GJ 179\\
        2457706.56513 &-4.417 &1.415 &CAR-VIS &GJ 179\\
        2459081.67004 &37.147 &1.322 &CAR-VIS &GJ 179\\
        2454807.68521 &-8564.622 &5.754 &HARPS-pre &GJ 179\\
        2454825.70109 &-8561.922 &2.131 &HARPS-pre &GJ 179\\
        2454826.67602 &-8569.180 &2.109 &HARPS-pre &GJ 179\\
        2454827.67585 &-8563.717 &1.898 &HARPS-pre &GJ 179\\
        2454828.70037 &-8560.871 &2.544 &HARPS-pre &GJ 179\\
         \hline
    \end{tabular}
    \begin{tablenotes}
    \item[1]  [1]The full table including all RVs of 19 stars used in this work is available in its entirety in machine-readable form in the online article.
    \end{tablenotes}
    \label{RVvalues}
\end{table*}

\section{Analysis}\label{jointfit}
Throughout the paper, we adopt a right-handed coordinate system defined by the orthonormal triad $[\hat{x}, \hat{y}, \hat{z}]$, where $\hat{x}$ points the direction of increasing R.A., $\hat{y}$ points the direction of increasing decl., and $\hat{z}$ points away from the observer (convention $\mathrm{III}$ of \citealt{PEXO2019}). This convention was also used in the astrometry models of Hipparcos and Gaia \citep{ESA1997, Lindegren2012}. 

We constrain the 3-D orbit of the cold Jupiters through a joint modeling of the RVs and Hipparcos-Gaia astrometric data, following the method developed by \cite{Feng2023} and subsequently modified by \cite{Xiao2024}, which has been widely used to search for and characterize wide-orbit companions (e.g., \citealt{Cui2026PNAS, Wuyiyun2026AJ, Pablo2026, Xiao2026ApJL}). 
The key fitting parameters in our model include seven orbital elements (orbital period $P$, RV semi-amplitude $K$, two parameters relevant to eccentricity and argument of periastron $\sqrt{e}\cos \omega$ and $\sqrt{e}\sin w$, orbital inclination $i$, longitude of ascending node $\Omega$, mean anomaly $M_{0}$ at epoch J2017.5 (JD2,457,936.875)), five astrometric offsets ($\Delta \alpha*, \Delta \delta, \Delta \varpi, \Delta \mu_{\alpha*}$ and $\Delta \mu_{\delta}$) of barycenter relative to GDR3, RV jitter terms $\sigma_{\rm jit}$ for each RV instrument, as well as an astrometric jitter $J_{\rm hip}$ for Hipparcos IAD, and an error inflation factor $S_{\rm gaia}$ for Gaia astrometry. 
The semi-major axis $a$ of the planet relative to the host, the mass of planet $M_{p}$, the planet-to-host mass ratio $q$, and the epoch of periastron passage $T_{p}$ are then derived from the above orbital elements. 

For the RV model, the likelihood function is expressed as
\begin{equation}
\begin{split}
  \mathcal{L}_{\rm RV}
  &=\prod\limits_{k=1}^{N_{\rm inst}}\prod\limits_{j=1}^{N_{{\rm RV},k}}
  \frac{1}{\sqrt{2\pi(\sigma_{j,k}^2+\sigma_{{\rm jit},k}^2)}} \times\\
  &\quad \exp\!\left( -\frac{(v_{j,k}-\hat{v}_{j,k}-\gamma_k)^2}
    {2(\sigma_{j,k}^2+\sigma_{{\rm jit},k}^2)} \right)~,
\end{split}
\end{equation}
where $N_{{\rm RV},k}$ is the number of measurements from instrument $k$, $N_{\rm RV}=\sum_k N_{{\rm RV},k}$, and $N_{\rm inst}$ is the number of instruments. 
$v_{j,k}$ and $\hat{v}_{j,k}$ are the observed and predicted RVs, respectively.
$\gamma_k$ denotes the RV offset for the $k$-th instrument, and $\sigma_{{\rm jit},k}$ represents the jitter term to account for unmodeled stellar or instrumental noise. 
We have marginalized out the RV zero-point following \citet{Brandt2021orvara}. 

The likelihood function for Hipparcos IAD can be calculated as
\begin{equation}
  \mathcal{L}_{\rm hip}=\prod\limits_{j=1}^{N_{\rm IAD}}\frac{1}{\sqrt{2\pi(\sigma_{j}^2+J_{\rm hip}^2)}}
  \times 
  \exp\!\left( -\frac{(\xi_j-\hat{\xi}_j)^2}{2(\sigma_{j}^2+J_{\rm hip}^2)} \right)~,
\end{equation}
where $N_{\rm IAD}$ is the number of epochs of Hipparcos IAD, $\sigma_j$ is the individual measurement uncertainty, and $J_{\rm hip}$ is the jitter term. 
Observed and predicted 1-D abscissae data are denoted with $\xi_j$ and $\hat{\xi}_j$, respectively. 
The abscissae $\hat{\xi}_j$ is synthesized by projecting the combined motion of the system's barycenter, the stellar reflex motion, and the parallax motion onto the 1-D scan direction of Hipparcos. 
At each Hipparcos IAD epoch, we propagate the barycentric astrometry from the reference catalog epoch to the Hipparcos epoch and compute its difference from the Hipparcos catalog astrometry. 
We add this difference to the modeled orbital reflex displacement, project the sum onto the Hipparcos scan direction, and compare it with the observed along-scan abscissa.

The likelihood function for GDR2 and GDR3 can be written as
\begin{equation}
\begin{split}
  \mathcal{L}_{\rm gaia}=
  &\prod\limits_{k=1}^{N_{\rm DR}}\frac{1}{\sqrt{(2\pi)^5|\Sigma_k(S_{\rm gaia}^2)|}}\times\\
  &\exp\left( -\frac{1}{2}(\Delta\vec{\iota}_k-\Delta\hat{\vec{\iota}}_k)^{T}[\Sigma_k(S_{\rm gaia}^2)]^{-1}(\Delta\vec{\iota}_k-\Delta\hat{\vec{\iota}}_k) \right)~,
\end{split}
\raisetag{-0.6\baselineskip}
\end{equation}
where $N_{\rm DR}$ represents the number of Gaia data releases ($N_{\rm DR}=2$ when we use both GDR2 and GDR3), $\Sigma_k$ is the covariance provided by Gaia catalog, and $S_{\rm gaia}$ is the Gaia error inflation factor. 
$\Delta\vec{\iota}_k$ is the five-parameter astrometric vector for catalog $k$ expressed relative to GDR3, e.g.,
\begin{equation}
\begin{split}
\Delta\vec{\iota}_k\equiv
&\,((\alpha_k -\alpha_{\rm DR3})\cos\delta_k,\,\delta_k-\delta_{\rm DR3},\,\\
&\varpi_k-\varpi_{\rm DR3},\,\mu_{\alpha k}-\mu_{\alpha \rm DR3},\,\mu_{\delta k}-\mu_{\delta \rm DR3}),
\end{split}
\end{equation}
where $\Delta\hat{\vec{\iota}}_k$ is the five synthetic astrometric vector obtained by fitting the standard five-parameter astrometric model to the simulated abscissae. 
For each Gaia data release, we evaluate the modeled along-scan positions at the GOST epochs included in that release, using the corresponding scan angles and parallax factors and applying the release-specific time baseline and valid-observation mask. 
We then fit the standard five-parameter astrometric model to these simulated along-scan positions.


We adopt uniform priors for most fitting parameters, with two exceptions: the orbital inclination $i$ is assigned an isotropic prior, equivalent to $\cos i\sim\mathcal{U}(-1,1)$, and the Gaia error inflation factor $S_{\rm Gaia}$ is assigned a truncated Gaussian prior, $S_{\rm Gaia}\sim\mathcal{TN}(1.0,\,0.1^2;\,1,\,2)$.
We adopt the stellar mass from Table~\ref{stellar_parameters} and use its Gaussian uncertainty when deriving $a$ and $M_p$.
With the joint likelihood, $\mathcal{L}=\mathcal{L}_{\rm RV}\cdot\mathcal{L}_{\rm hip}\cdot\mathcal{L}_{\rm gaia}$, we derive the orbital solution by sampling the posterior via the parallel-tempering Markov Chain Monte Carlo (MCMC) sampler \texttt{ptemcee} \citep{Vousden2016}. 
We employ 30 temperatures, 100 walkers, and 80,000 steps per chain to generate posterior distributions for all the fitting parameters, with the first 40,000 steps being discarded as burn-in. 
Among the 19 systems in the sample, five of them including GJ 317, GJ 3512, GJ 9066, TOI-756 and TYC 2187-512-1 do not have Hipparcos measurements. We therefore constrain their orbital solutions using Gaia astrometry, i.e., $\mathcal{L}=\mathcal{L}_{\rm RV}\cdot\mathcal{L}_{\rm gaia}$.

\section{Results}\label{results}

\begin{figure*}[htb]
\centering
\includegraphics[width=0.99\textwidth]{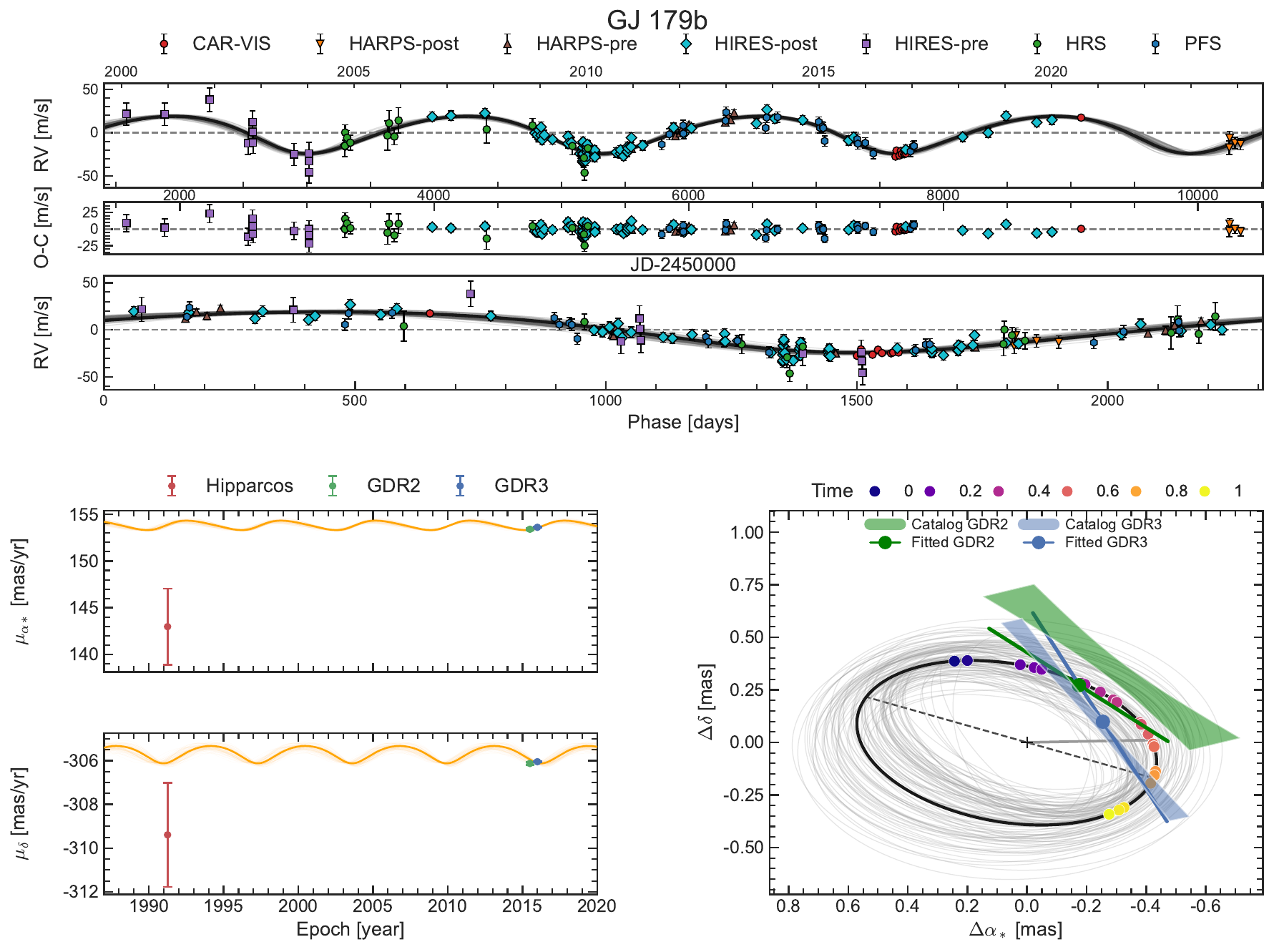}
\caption{RV + Astrometry fitting results for GJ 179 from the joint analysis. 
\textbf{(a)} Top panels: RV fit for GJ 179, including the time series of RVs (top), residuals  (middle) and the phase-folded RVs (bottom) along with the best-fit Keplerian model shown as the black curve. 
\textbf{(b)} Bottom-left panels: Proper-motion evolution of GJ 179 predicted by the joint fit. 
The upper and lower panels show the right-ascension and declination components, $\mu_{\alpha *}$ and $\mu_\delta$, respectively.
The thick purple or orange curves show the instantaneous model proper motion for the two inclination branches, $i_{b} < 90^\circ$ or $i_{b} > 90^\circ$, respectively, both of which include the fitted barycentric proper motion and time-dependent reflex proper motion induced by GJ 179b. 
Thin curves represent orbital solutions randomly selected from the MCMC posterior distributions.  
The red, green, and blue points denote the Hipparcos, GDR2, and GDR3 catalog proper motions with their uncertainties. 
\textbf{(c)} Bottom-right panel: Astrometric reflex orbit of GJ 179 induced by GJ 179b. 
The plus sign marks the system barycenter.
The thick black curve shows the best-fit orbit, while thin gray curves show 100 posterior draws. 
The green and blue shaded regions indicate the barycentric-motion-subtracted uncertainty envelopes for GDR2 and GDR3.
The green and blue circles mark companion-induced position offsets at the GDR2 and GDR3 reference epochs, and the associated solid lines show their evolution due to the fitted proper-motion offsets. 
The colored points along the black orbit show the model-predicted astrometric positions at the Gaia scanning epochs and are color-coded by normalized observing time, increasing from purple to yellow. }
\label{fig:fit_GJ179}
\end{figure*}

For each candidate, we generate a diagnostic plot as in Figure \ref{fig:fit_GJ179} together with a posterior plot as in Figure \ref{fig:posterior_GJ179}. 
We define a robust orbital solution if both the upper and lower uncertainties on the orbital inclination are smaller than $30^{\circ}$. 
This threshold is more stringent than the half-width ($\simeq 43^\circ$) of the central 68\% interval of the isotropic inclination prior, $\cos i\sim\mathcal{U}(-1,1)$, and is adopted as an operational criterion for identifying systems whose inclinations are meaningfully constrained by the data.
Overall, we measure the robust absolute planet masses of 13 cold Jupiters in 11 M dwarf systems including GJ 179b (Figure \ref{fig:fit_GJ179}), GJ 317b (Figure \ref{fig:fit_GJ317}), GJ 328b (Figure \ref{fig:fit_GJ328}), GJ 463b (Figure \ref{fig:fit_GJ463}), GJ 508.2b (Figure \ref{fig:fit_GJ508.2}), GJ 649b (Figure \ref{fig:fit_GJ649}), GJ 676Ab and c (Figure \ref{fig:fit_GJ676A}), GJ832b (Figure \ref{fig:fit_GJ832}), GJ 849b and c (Figure \ref{fig:fit_GJ849}), GJ 9066c (Figure \ref{fig:fit_GJ9066}) as well as HIP 88962b (Figure \ref{fig:fit_HIP88962}). Except for planetary systems, we also find that the RV-detected cold Jupiter orbiting HIP 79431 is a brown dwarf companion around an M star with a nearly face-on orbit (Figure \ref{fig:fit_HIP79431}). The results of GJ 317c, GJ 876b and c, GJ 1148b and c, GJ 3512b and c, GJ 9773b, HIP 12961b, TOI-756c and TYC 2187-512-1b are ambiguous given their large inclination uncertainties hence we do not include them. We list the key results of systems with robust solutions in Table~\ref{key_result_summary}, and discuss each of them in detail in the following sub-sections. We list all posteriors of every robust system in Table~\ref{sample_result_summary}. 

\begin{table*}[htb]
    \centering
    {\renewcommand{\arraystretch}{1.25}
    \caption{Key parameters of 13 cold Jupiters and one brown dwarf around M dwarfs with robust orbital solutions}
    \begin{tabular}{ccccccc}
        \hline\hline
        Target       &Planet Mass &Orbital Period &Semi-major axis &Inclination &Eccentricity &Mass ratio \\
        &[$M_{\rm Jup}$] &[days] &[AU] &[deg] & & \\\hline
        GJ 179b & $0.982^{+0.156}_{-0.128}$ & $2318.5^{+26.3}_{-23.9}$ & $2.449^{+0.048}_{-0.048}$ & $133.0^{+7.8}_{-9.9}$ & $0.147^{+0.033}_{-0.033}$ & $0.0026^{+0.0004}_{-0.0003}$ \\
        GJ 317b$^{[1]}$ & $2.855^{+0.513}_{-0.477}$ & $695.0^{+0.5}_{-0.3}$ & $1.156^{+0.014}_{-0.029}$ & $37.8^{+10.3}_{-6.9}$ & $0.078^{+0.007}_{-0.009}$ & $0.0065^{+0.0014}_{-0.0011}$ \\
          & $2.316^{+0.329}_{-0.283}$ & $695.0^{+0.3}_{-0.3}$ & $1.154^{+0.019}_{-0.019}$ & $130.7^{+7.4}_{-10.1}$ & $0.079^{+0.008}_{-0.008}$ & $0.0052^{+0.0007}_{-0.0006}$ \\
        GJ 328b & $2.310^{+0.155}_{-0.129}$ & $3766.8^{+17.6}_{-17.7}$ & $4.062^{+0.045}_{-0.045}$ & $96.9^{+13.0}_{-15.3}$ & $0.228^{+0.015}_{-0.015}$ & $0.0035^{+0.0002}_{-0.0002}$ \\
        GJ 463b$^{[2]}$ & $3.178^{+0.279}_{-0.274}$ & $3464.6^{+21.8}_{-20.9}$ & $3.496^{+0.051}_{-0.051}$ & $31.1^{+3.2}_{-2.7}$ & $0.138^{+0.016}_{-0.016}$ & $0.0064^{+0.0005}_{-0.0005}$ \\
        & $2.919^{+0.253}_{-0.255}$ & $3463.6^{+20.8}_{-19.5}$ & $3.496^{+0.052}_{-0.051}$ & $145.8^{+3.0}_{-3.7}$ & $0.138^{+0.016}_{-0.016}$ & $0.0059^{+0.0005}_{-0.0005}$ \\
        GJ 508.2b & $5.835^{+1.129}_{-1.090}$ & $8591.8^{+1333.3}_{-1259.4}$ & $6.661^{+0.869}_{-0.825}$ & $19.2^{+3.4}_{-2.0}$ & $0.530^{+0.050}_{-0.059}$ & $0.0105^{+0.0015}_{-0.0015}$ \\
        GJ 649b & $0.835^{+0.223}_{-0.212}$ & $600.3^{+1.5}_{-1.5}$ & $1.116^{+0.014}_{-0.015}$ & $18.4^{+7.4}_{-4.0}$ & $0.118^{+0.040}_{-0.040}$ & $0.0016^{+0.0004}_{-0.0004}$ \\
        GJ 676Ab & $5.353^{+0.244}_{-0.232}$ & $1052.9^{+0.6}_{-0.6}$ & $1.737^{+0.019}_{-0.019}$ & $125.1^{+3.0}_{-3.3}$ & $0.316^{+0.004}_{-0.004}$ & $0.0082^{+0.0003}_{-0.0003}$ \\
        GJ 676Ac & $9.254^{+0.358}_{-0.360}$ & $11038.6^{+1059.0}_{-720.6}$ & $8.361^{+0.532}_{-0.377}$ & $113.3^{+3.7}_{-4.6}$ & $0.137^{+0.047}_{-0.037}$ & $0.0141^{+0.0005}_{-0.0005}$ \\
        GJ 832b & $0.938^{+0.038}_{-0.038}$ & $3736.7^{+13.1}_{-13.2}$ & $3.580^{+0.053}_{-0.056}$ & $54.1^{+2.1}_{-2.0}$ & $0.013^{+0.013}_{-0.009}$ & $0.0020^{+0.0001}_{-0.0001}$ \\
        GJ 849b & $0.929^{+0.056}_{-0.037}$ & $1921.6^{+4.3}_{-4.3}$ & $2.338^{+0.034}_{-0.035}$ & $89.7^{+13.5}_{-16.0}$ & $0.048^{+0.012}_{-0.012}$ & $0.0019^{+0.0001}_{-0.0001}$ \\
        GJ 849c & $0.947^{+0.086}_{-0.054}$ & $5873.0^{+93.4}_{-86.6}$ & $4.926^{+0.090}_{-0.087}$ & $93.3^{+17.3}_{-19.6}$ & $0.058^{+0.024}_{-0.022}$ & $0.0020^{+0.0002}_{-0.0001}$ \\
        GJ 9066c$^{[3]}$ & $0.296^{+0.128}_{-0.075}$ & $769.4^{+3.9}_{-2.9}$ & $0.875^{+0.035}_{-0.040}$ & $46.5^{+28.0}_{-16.2}$ & $0.467^{+0.070}_{-0.076}$ & $0.0019^{+0.0008}_{-0.0005}$ \\
            & $0.404^{+0.169}_{-0.129}$ & $769.4^{+3.8}_{-3.0}$ & $0.874^{+0.037}_{-0.041}$ & $148.6^{+9.9}_{-18.0}$ & $0.461^{+0.067}_{-0.075}$ & $0.0026^{+0.0011}_{-0.0008}$ \\
        HIP 79431B$^{[4]}$ & $31.321^{+8.459}_{-9.653}$ & $109.919^{+0.008}_{-0.008}$ & $0.358^{+0.005}_{-0.006}$ & $176.3^{+0.8}_{-1.6}$ & $0.332^{+0.010}_{-0.010}$ & $0.0628^{+0.0169}_{-0.0194}$ \\
        HIP 88962b & $4.830^{+0.630}_{-0.517}$ & $7054.9^{+693.3}_{-482.3}$ & $6.154^{+0.461}_{-0.396}$ & $90.7^{+22.1}_{-23.1}$ & $0.718^{+0.032}_{-0.029}$ & $0.0074^{+0.0008}_{-0.0005}$ \\
         \hline
    \label{key_result_summary}
    \end{tabular}}
    \begin{tablenotes}
       \item[1]  [1]\ $\Delta$BIC between two solutions of GJ 317b is ${\rm BIC}_{i_b>90^{\circ}}-{\rm BIC}_{i_b<90^{\circ}}=-6.709$, and the retrograde-orbit ($i_b>90^{\circ}$) solution is slightly preferred. 
       [2]\ $\Delta$BIC between two solutions of GJ 463b is ${\rm BIC}_{i_b>90^{\circ}}-{\rm BIC}_{i_b<90^{\circ}}=1.57$, and the prograde-orbit solution ($i_b<90^{\circ}$) is slightly preferred. 
       [3]\ $\Delta$BIC between two solutions of GJ 9066c is ${\rm BIC}_{i_c>90^{\circ}}-{\rm BIC}_{i_c<90^{\circ}}=-5.24$, and the retrograde-orbit solution ($i_c>90^{\circ}$) is slightly preferred. 
       [4]\ HIP 79431B is a brown dwarf companion. 
    \end{tablenotes}
\end{table*}

\subsection{GJ 179b}

GJ 179b was first discovered by \cite{Howard2010} with a period of $6.26\pm0.16$ yrs and a minimum planet mass of $M_p\sin i=0.82\pm0.07~M_J$. Through the California Legacy Survey, \cite{Rosenthal2021} revisited the system and updated the measurements to $P=2333\pm36$~days $M_p\sin i=0.752\pm0.041~M_J$. More recently, \cite{Xiao2023} carried out full orbit fits combining RVs from HARPS, HIRES and HRS, and astrometric data from Hipparcos and GDR3, reporting an absolute mass of $M_p=0.95^{+0.16}_{-0.11}~M_J$ and an inclination of $61^{+16}_{-13}$$^{\circ}$. 
Here, we include RV measurements from CARMENES and PFS, and perform a joint analysis together with Hipparcos, GDR2 and GDR3, yielding a refined mass measurement of $M_p=0.98^{+0.16}_{-0.13}~M_J$ and an inclination of $133.0^{+7.8}_{-9.9}$$^{\circ}$. 

\subsection{GJ 317b}

\cite{Johnson2007GJ317} first reported the discovery of GJ 317b, a giant planet with a minimum mass of $M_p\sin i\sim1.2~M_J$ on a 693-day eccentric ($e=0.19\pm0.06$) orbit. Subsequent works from \cite{Feng2020}, \cite{Rosenthal2021} and \cite{Morales2025} refined the minimum mass measurements to $1.753\pm0.058~M_J$, $1.853\pm0.037~M_J$ and $1.688\pm0.041~M_J$, respectively. In particular, \cite{Anglada2012GJ317} obtained 18 astrometric observations spanning 2.5 years, and measured a true planet mass of $2.5^{+0.7}_{-0.4}~M_J$ and an inclination of $45^{+30}_{-10}$$^{\circ}$. 
Our joint analysis yields two degenerate solutions with the retrograde-orbit solution ($M_p=2.32^{+0.33}_{-0.28}~M_J$, $i=130.7^{+7.4}_{-10.1}$$^{\circ}$) preferred with a Bayesian Information Criterion (BIC) improvement of 6.709. Note that GJ 317 contains another wide-orbit companion GJ 317c \citep{Anglada2012GJ317,Bryan2016,Feng2020,Rosenthal2021,Morales2025}. Our fit gives an inclination uncertainty exceeding 30 degrees, which we regard as non-robust result. Therefore, we exclude this planet from the final results.

\subsection{GJ 328b}

GJ 328b was announced by \cite{Robertson2013} using RV data from HIRES, HRS and Tull, covering a baseline of about 10 years. The planet was found to have an orbital period of $4100\pm300$~days and a minimum mass of $M_p\sin i=2.30\pm0.13~M_J$. Recent work from \cite{Pinamonti2023} collected public RV measurements from SOPHIE and obtained 131 HARPS-N RVs over a time span of 1132 days. The authors determined a minimum mass of $2.51\pm0.23~M_J$ and an orbital period of $3771\pm17$~days for GJ 328b. In addition, they identified a new inner small planet GJ 328c with $P=241.8^{+1.3}_{-1.7}$~days and $M_p\sin i=21.4^{+3.4}_{-3.2}~M_\oplus$. 
Since the astrometric signature of GJ 328c is negligible, we do not consider this inner planet GJ 328c in the joint fit. After including the public HARPS-post data taken between UT 2024 January 12 and 2024 April 28, we find that GJ 328b has a mass of $M_p=2.31^{+0.16}_{-0.13}~M_J$. Our mass measurement is slightly smaller than the minimum mass derived by \cite{Pinamonti2023}, mainly because we adopt a smaller stellar mass $0.63\pm0.02$ from TIC than the value $0.65\pm0.08~M_\odot$ used in \cite{Pinamonti2023}.

\subsection{GJ 463b}

Based on the RVs from HIRES and HRS over 10 years, \cite{Endl2022} reported that GJ 463b is a giant planet companion with an orbital period of $3448^{+110}_{-88}$~days and a minimum mass of $1.55\pm0.15~M_J$, which was later updated to $1.649^{+0.044}_{-0.045}~M_J$ after including the CARMENES data collected by \cite{Morales2025}. Both \cite{Sozzetti2023} and \cite{Piccinini2026} carried out a joint fit with Hipparcos, GDR3 astrometry together with HRS and HIRES RVs, leading to two degenerate solutions: one prograde-orbit solution ($M_p=3.4\pm0.3~M_J$ and $27\pm3$$^{\circ}$ from \citealt{Sozzetti2023}; $M_p=3.3^{+0.4}_{-0.3}~M_J$ and $27.7^{+4.6}_{-3.9}$$^{\circ}$ from \citealt{Piccinini2026}) and one retrograde-orbit solution ($M_p=3.6\pm0.4~M_J$ and $152^{+2}_{-3}$$^{\circ}$ from \citealt{Sozzetti2023}; $M_p=3.4\pm0.4~M_J$ and $152.3^{+3.9}_{-4.6}$$^{\circ}$ from \citealt{Piccinini2026}). 
In this work, we include not only the CARMENES RV data from \cite{Morales2025} and 5 additional measurements but also the GDR2 astrometry in the fit. 
However, similar as \cite{Sozzetti2023} and \cite{Piccinini2026}, our analysis cannot break the orbital degeneracy. We find that the prograde-orbit solution ($M_p=3.18^{+0.28}_{-0.27}~M_J$ and $31.1^{+3.2}_{-2.7}$$^{\circ}$) is slightly preferred with a BIC improvement of 1.57 in contrast to the retrograde orbit. 

\subsection{GJ 508.2b}

GJ 508.2b was recently confirmed by \cite{Morales2025} with CARMENES observations. Owing to the limited RV phase coverage, the authors only reported lower limits on the orbit period ($P>5300$~days), minimum mass ($M_p\sin i>1.85~M_J$) and orbital eccentricity ($e>0.27$). By assuming the lower limit period, \cite{Morales2025} derived a planet mass of $M_p=6.0^{+1.7}_{-0.2}~M_J$. After including six new CARMENES RVs, our joint analysis constrains the orbital period of GJ 508.2b to be $8591.8^{+1333.3}_{-1259.4}$~days. Although the orbital period remains poorly determined, the absolute mass, inclination and eccentricity of GJ 508.2b are measured to be $M_p=5.84^{+1.13}_{-1.09}~M_J$, $i=19.2^{+3.4}_{-2.0}$$^{\circ}$, and $e=0.530^{+0.050}_{-0.059}$, respectively. 

\subsection{GJ 649b}

\cite{Johnson2010} reported that GJ 649b has a minimum mass of $0.328\pm0.032 \ M_J$ and a period of $598.3\pm4.2$~days based on Keck/HIRES data. With additional HIRES RVs collected through the California Legacy Survey, \cite{Rosenthal2021} refined the orbital period to $P=604.8^{+3.4}_{-3.5}$~days and the minimum planet mass to $M_p\sin i=0.275^{+0.022}_{-0.021}~M_J$. More recently, \cite{Pinamonti2023} retrieved all HIRES data as well as CARMENES RV measurements, and obtained updated values of $P=600.1\pm1.7$~days and $M_p\sin i=0.258^{+0.023}_{-0.022}~M_J$. Using the same RV data as \cite{Pinamonti2023}, our RV+Astrometry fit yields an absolute planet mass of $0.84^{+0.22}_{-0.21}~M_J$ and an inclination of $18.4^{+7.4}_{-4.0}$$^{\circ}$.

\subsection{GJ 676Ab and GJ 676Ac}

GJ 676 is a binary star system \citep{Forveille2011}, and the primary star GJ 676A hosts four confirmed planets \citep{Anglada2012GJ676,Sahlmann2016,Feng2022}. Two small inner planets GJ 676Ad and GJ 676Ae have orbital periods of 3.6~days and 35.4 days, both of which lie beyond the detection limit of the astrometric method. The two outer planets are both gas giants. GJ 676Ab has a minimum mass of about $5~M_J$ and a period of about 1050 days \citep{Forveille2011,Anglada2012GJ676}. \cite{Sahlmann2016} obtained position measurements, and constrained the inclination to $45^{+21}_{-11}$$^{\circ}$ and the mass to $M_p=6.7^{+1.8}_{-1.5}~M_J$. Similar results of $48.9^{+3.3}_{-2.8}$$^{\circ}$ and $M_p=5.792^{+0.469}_{-0.477}~M_J$ were obtained by \cite{Feng2022}. Regarding GJ 676Ac, \cite{Sahlmann2016} reported an orbital period of about 7340 days and a minimum mass of $M_p\sin i=6.8\pm 0.1~M_J$, assuming a circular orbit. However, \cite{Feng2022} found the orbit is in fact eccentric ($e=0.295^{+0.033}_{-0.049}$). They derived a much longer period $P=13921.4^{+1232.3}_{-1514.6}$~days with an absolute companion mass of $13.492^{+1.046}_{-1.127}~M_J$ and an inclination of $33.7^{+1.4}_{-1.3}$$^{\circ}$, indicating that the companion is likely a brown dwarf. After including new HARPS-post data collected after 2018 and PFS data obtained after 2021 in the joint analysis, we confirm that GJ 676Ac is a massive giant planet with an absolute mass of $9.25\pm0.36~M_J$. Its orbit is determined to have a period of $11038.6^{+1059.0}_{-720.6}$~days and an eccentricity of $0.137^{+0.047}_{-0.037}$, inclined by $113.3^{+3.7}_{-4.6}$$^{\circ}$. We also refine the period, mass, inclination of GJ 676Ab to be $P=1052.9^{+0.56}_{-0.57}$~days, $M_p=5.35^{+0.24}_{-0.23}~M_J$ and $i=125.1^{+3.0}_{-3.3}$$^{\circ}$.

\subsection{GJ 832b}

GJ 832 hosts a giant planet with an orbital period of about 10 years with a minimum mass of $0.68\pm0.09~M_J$ \citep{Bailey2009,Bonfils2013}. In addition, \cite{Wittenmyer2014} reported an inner planet GJ 832b, but the signal was confirmed as a false positive by \cite{Gorrini2022}. Later work by \cite{Xiao2023} and \cite{Philipot2023} both conducted similar joint fits of RVs together with Hipparcos and GDR3 astrometric acceleration, leading to absolute mass measurements of $0.80^{+0.12}_{-0.11}~M_J$ and $0.990^{+0.090}_{-0.080}~M_J$ as well as inclinations of $54.9^{+6.6}_{-4.9}$$^{\circ}$ and $51\pm3$$^{\circ}$ for GJ 832b. Recently, \cite{An2025} repeated the analysis and reported a similar mass measurement $M_p=1.039^{+0.044}_{-0.043}~M_J$ but a different inclination of $131.6^{+2.7}_{-3.4}$$^{\circ}$. The results from our analysis, $M_p=0.94\pm0.04~M_J$ and $i=54.1^{+2.1}_{-2.0}$$^{\circ}$, are more consistent with the findings from \cite{Xiao2023} and \cite{Philipot2023}. 

\subsection{GJ 849b and GJ 849c}

GJ 849 hosts two confirmed cold Jupiters. The inner giant planet, GJ 849b, has a minimum mass of about $0.9~M_J$ and a period of about 1900 days \citep{Butler2006catalog,Bonfils2013,Montet2014,Feng2015,Rosenthal2021,Pinamonti2023}. The outer planet GJ 849c was first reported by \cite{Montet2014}, with a highly uncertain period of $7068^{+6282}_{-2166}$~days and a minimum mass of $M_p\sin i=0.702^{+0.344}_{-0.203}~M_J$. \cite{Rosenthal2021} and \cite{Pinamonti2023} improve the orbital solution by including HIRES and HARPS-N RVs. \cite{Rosenthal2021} reported an orbital period of $5750^{+130}_{-120}$~days and a minimum mass of $M_p\sin i=1.079\pm0.053~M_J$, while \cite{Pinamonti2023} found a period of $5990^{+110}_{-100}$~days and a minimum mass of $M_p\sin i=0.99\pm0.11~M_J$. We find that GJ 849b has a true mass of $0.93^{+0.06}_{-0.04}~M_J$ with $i=89.7^{+13.5}_{-16.0}$$^{\circ}$, and GJ 849c has a similar mass of $0.95^{+0.09}_{-0.05}~M_J$ with $i=93.3^{+17.3}_{-19.6}$$^{\circ}$.

\subsection{GJ 9066c}

GJ 9066 was initially reported to host two planets \citep{Feng2020}, but the inner cold Neptune GJ 9066b was ruled out as a false positive caused by stellar activity \citep{Quirrenbach2022}. In terms of GJ 9066c, \cite{Feng2020} and \cite{Quirrenbach2022} both measured a minimum mass of about $0.2~M_J$. Through the joint analysis, we confirm that GJ 9066c is a giant planet. However, we find two degenerate solutions: a prograde-orbit solution ($M_p=0.30^{+0.13}_{-0.08}~M_J$ and $i=46.5^{+28.0}_{-16.2}$$^{\circ}$) and a retrograde-orbit solution ($M_p=0.40^{+0.17}_{-0.13}~M_J$ and $i=148.6^{+9.9}_{-18.0}$$^{\circ}$). The retrograde-orbit solution is preferred with a BIC improvement of 5.24. 

\subsection{HIP 79431B}

\cite{Apps2010} first reported the planet orbiting HIP 79431, with a minimum mass of about $2.1~M_J$, an orbital period of $111.7\pm0.7$~days, and an eccentricity of $0.29\pm0.02$. 
Our fit identifies the companion as a brown dwarf, with a true mass of $31.32^{+8.46}_{-9.65}~M_J$ on a nearly face-on orbit ($i=176.3^{+0.7}_{-1.6}$$^{\circ}$). 
The fitted Hipparcos astrometric jitter is relatively large, $J_{\rm hip}=22.17^{+3.17}_{-2.62}$ mas, indicating substantial excess scatter in the Hipparcos IAD under our adopted noise model and consequently reducing their weight in the joint fit. 
The inferred mass and inclination are therefore likely driven primarily by the GDR2 and GDR3 catalog astrometry.
Using the Gaia proper-motion anomaly and the excess noise, recent work from \cite{Destriez2026} likewise argued that HIP 79431 is likely a highly-inclined system and may host two companions.

\subsection{HIP 88962b}

HIP 88962b was recently confirmed as an eccentric cold Jupiter through PFS observations \citep{Boss2026}. Our joint analysis reveals that the planet has an orbital period of $7054.9^{+693.3}_{-482.3}$~days, a mass of $M_p=4.83^{+0.63}_{-0.52}~M_J$ and an eccentricity of $e=0.718^{+0.032}_{-0.029}$. The orbit inclination is determined to be $i=90.7^{+22.1}_{-23.1}$$^{\circ}$. 

\section{Discussion}\label{discussion}

\subsection{Mass Ratios of Giant Planets Orbiting M Dwarfs}

\begin{figure*}
    \centering
    \includegraphics[width=0.99\textwidth]{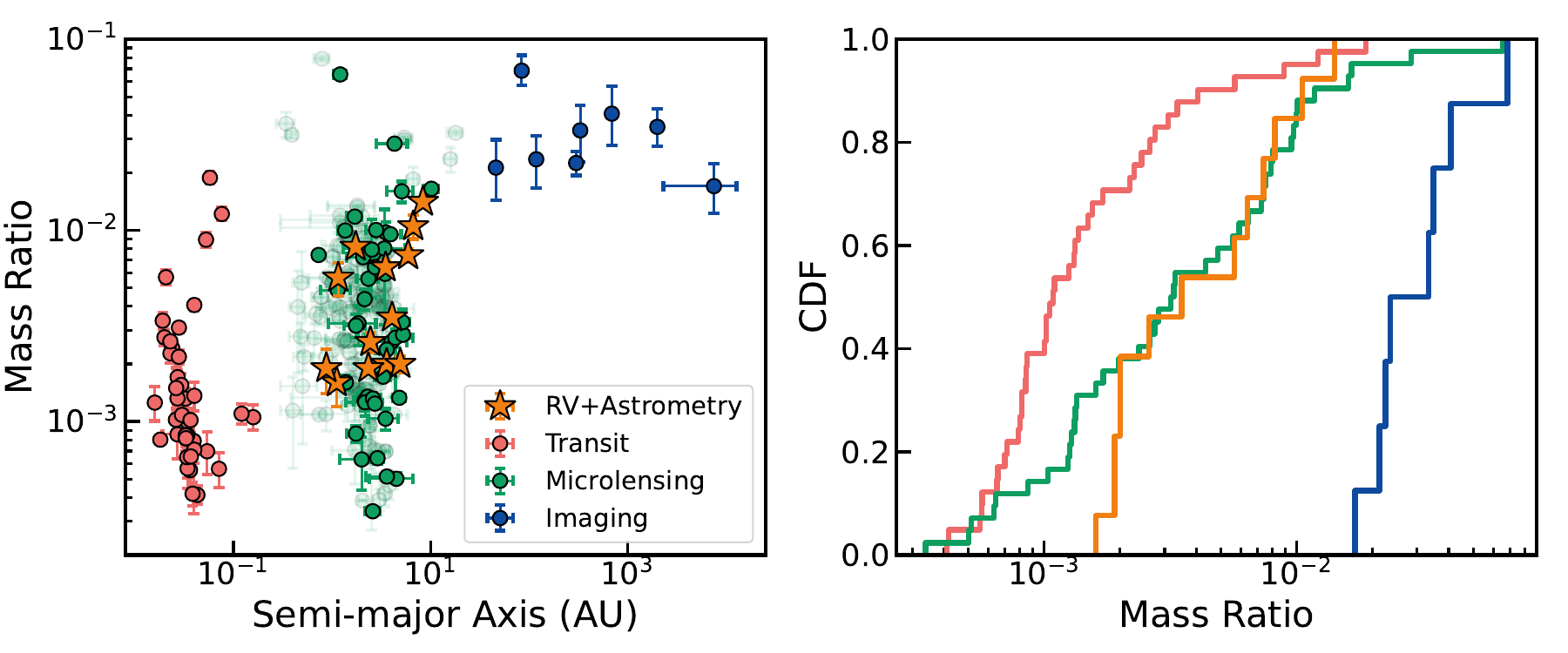}
    \caption{Left: The mass ratio and semi-major axis distributions of all confirmed giant planets around M dwarfs with absolute planet mass measurements, colored by different detection and characterization techniques. The 13 cold Jupiters studied in this work are marked as orange stars. The transparent green dots are the cold Jupiters found by microlensing but with $\geq 2$ degenerate orbital solutions, which are excluded from the analysis. Right: The cumulative mass ratio distribution function of giant planets around M dwarfs. There is a possible paucity of massive hot Jupiters around M stars. Without considering detection bias, cold Jupiters around M dwarfs characterized by microlensing and RV+Astrometry exhibit similar mass ratio distributions.}
    \label{fig:mass_ratio_semi_major_axis}
\end{figure*}

In the left panel of Figure~\ref{fig:mass_ratio_semi_major_axis}, we present the planet-to-star mass ratio as a function of semi-major axis for all confirmed giant planets orbiting M dwarfs with absolute planet mass measurements. Among 41 transiting systems (red dots), the majority of them have semi-major axes within 0.1 AU, except for TOI-1899b \citep{Canas2020} and K2-419Ab \citep{Kanodia2024}. In addition, there are 8 directly imaged cold Jupiters around M stars (blue dots) with semi-major axes beyond 10 AU \citep{Ducourant2008,Ireland2011,Patience2012,Delorme2013,Naud2014,Zhang2021,Gaidos2022,Walker2026}, and all planets have masses greater than $4~M_J$, which is due to the observational bias. Within the semi-major axis range where transit and imaging methods are sensitive, microlensing has detected numerous wide-orbit cold Jupiters around M dwarfs with $1 \lesssim a \lesssim 10$~AU \citep{Akeson2013,Christiansen2025}. However, only 42 systems were reported to have unique orbital solutions (green dots). We excluded others with at least two degenerate orbital solutions (transparent green dots) from the analysis below. Finally, the semi-major axes of 13 cold Jupiters orbiting M dwarfs characterized in this work (orange stars) through RV+Astrometry analysis overlap with those of the microlensing sample \footnote{For the microlensing sample, we consider only the projected semi-major axis $a_{p}$ here. To account for random geometry orientations in the sky, one needs to multiply by a factor of $4/\pi$ to estimate the true semi-major axis (i.e., $a=a_p\times (4/\pi)$). However, this correction is valid only in a statistical sense and does not hold for individual microlensing planetary systems.}.

The right panel of Figure~\ref{fig:mass_ratio_semi_major_axis} shows the cumulative distribution functions of the mass ratios of giant planets orbiting M dwarfs, colored by detection techniques. We find a potential dearth of high-mass-ratio hot Jupiters around M dwarfs \citep[see also][]{Gantoi530}. Most transiting hot Jupiters orbiting M stars have mass ratios around $10^{-3}$, and only five of them (TOI-4201b, \citealt{Gan2023TOI4201,Hartman2023,Delamer2024}; TIC 46432937b, \citealt{Hartman2024}; TOI-2379b, \citealt{Bryant2024}; TOI-6303b and TOI-6330b, \citealt{Hotnisky2024}) have mass ratios above $4\times10^{-3}$. Meanwhile, it turns out that there is probably a large population of high-mass-ratio cold Jupiter systems with $q>4\times10^{-3}$ according to the microlensing and our RV+Astrometry results. We carry out two-sample Kolmogorov–Smirnov tests between the transit and microlensing samples, as well as between the transit and the RV+Astrometry samples using \code{SciPy} \citep{Virtanen2020}, yielding p-values of $2.9\times10^{-3}$ and $4.9\times10^{-5}$. Therefore, we conclude that the null hypothesis, the mass ratio distributions of hot and cold Jupiters are the same, could be rejected at the 3-4$\sigma$ confidence level. The lack of massive hot Jupiters around M stars is likely physical. 
The detection sensitivity of the transit method depend primarily on the planet radius rather than its mass. 
Since the radii of Jovian planets are roughly concentrated around $1\,R_{\rm J}$ over a wide mass range of $0.2$–$13\,M_{\rm J}$ \citep{Hatzes2015,Chen2017,Bashi2017,Edmondson2023,Muller2024}, the transit sample is not expected to be strongly biased in mass ratio over this range and may approximately reflect the underlying distribution. 
This argument is particularly applicable to gas giants around M dwarfs that receive low incident flux, for which irradiation-driven radius inflation is expected to be weak. 
The higher mass ratios of cold Jupiters relative to hot and warm Jupiters around M dwarfs may therefore reflect an intrinsic difference, although other selection effects cannot be ruled out.

Additionally, cold Jupiters around M dwarfs characterized by RV+Astrometry here and those from microlensing exhibit analogous mass ratio distributions, despite two populations residing in distinct regions of the Milky Way. Planets found by microlensing are mostly located in the Galactic bulge, a dense stellar field populated by numerous old stars \citep{Mao1991}. Such an environment differs from that of astrometric giant planets, which are around younger stars in the local Galactic thin disk, where the stellar number density is much lower \citep{Adibekyan2013}. While the similarity of mass ratio distributions hints that the Galactic environment may not have a notable effect on the properties of cold Jupiters around M dwarfs with semi-major axis between 1 and 10 AU, the sample of astrometric giant planets around M dwarfs is highly limited and heterogeneous. Furthermore, the uncorrected detection completeness of two samples makes it difficult to draw a firm conclusion. The upcoming Gaia DR4 is expected to bring the detections of thousands of cold Jupiters orbiting local M stars \citep{Perryman2014,Lammers2026}, which will provide a golden clean sample to compare with the transit and microlensing findings.

\subsection{Cold Jupiters Orbiting M Dwarfs and More Massive Stars}

\begin{figure}
    \centering
    \includegraphics[width=0.49\textwidth]{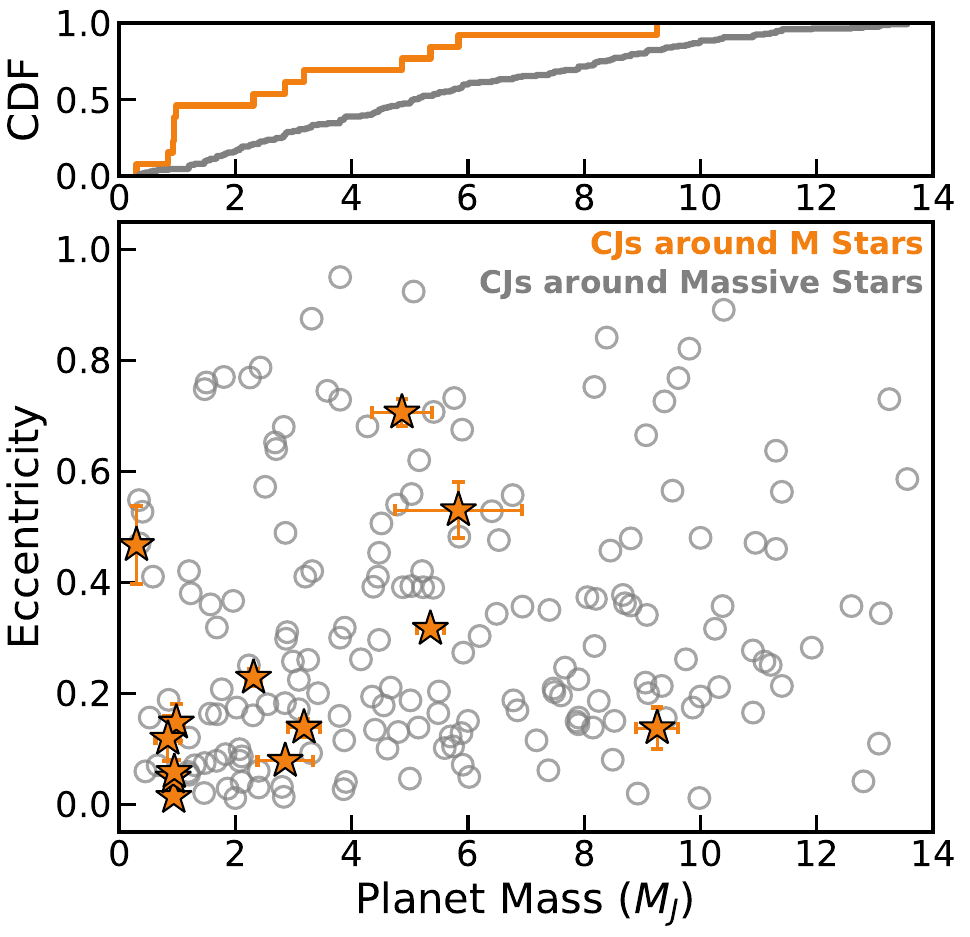}
    \caption{The eccentricity and absolute planet mass distributions of 13 cold Jupiters orbiting M (orange, $M_\ast \leq 0.65~M_\odot$) characterized in this work as well as 177 cold Jupiters around more massive (gray, $M_\ast > 0.65~M_\odot$) stars from NASA Exoplanet Archive \citep{Akeson2013,Christiansen2025} with $0.2<M_p<13.6~M_J$ and semi-major axis $a>0.5$~AU. The cumulative distribution functions of planet mass are shown in the top panel.}
    \label{fig:ecc_semi_major_axis}
\end{figure}

Figure~\ref{fig:ecc_semi_major_axis} illustrates the eccentricity and absolute planet mass distributions of 13 cold Jupiters around M stars (orange) characterized in this work with $0.2<M_p<13.6~M_J$ and semi-major axes greater than 0.5~AU, together with 177 cold Jupiters orbiting more massive stars ($M_\ast>0.65~M_\odot$; gray) stars that meet the same conditions. The two samples exhibit no clear difference in their eccentricity distributions. Although detection sensitivity also plays a role as the star’s astrometric displacement is proportional to the mass ratio of the system, we find that, on average, cold Jupiters around massive stars tend to be also more massive with a median planet mass of $5.1~M_J$, compared with cold Jupiters orbiting M stars that have a median planet mass of $2.3~M_J$. Specifically, 7 out of 13 (53.8\%) cold Jupiters orbiting M stars have absolute masses above $1~M_J$, whereas this fraction rises to 169 out of 177 (95.5\%) for cold Jupiters orbiting more massive stars. We perform a two-sample Kolmogorov–Smirnov tests for the masses of cold Jupiters around M and more massive stars, which leads to a p-value of 0.02, indicating that two samples are not from the same distribution. 

Such a feature might be an outcome of the core accretion formation \citep{Pollack1996}. Since M dwarfs have lower-mass protoplanetary disks \citep{Pascucci2016,Ansdell2016,Manara2018}, less gas can be accreted before disk dissipation, leading to lower planet masses. Furthermore, there are probably multiple giant planets formed around massive stars thanks to the massive disks that can provide abundant solid materials to form multiple compact cores and initiate runaway accretion \citep{Burn2021}. Through head-on collision \citep{Li2010,Liu2015}, two low-mass gas giants can merge to form a massive planet, which may account for the higher mass of cold Jupiters around massive stars compared to M dwarfs. \cite{Pan2023} claimed that the planet merger channel can also produce giant planets around the late-M dwarfs with mass about $0.1\sim0.2~M_\odot$, for example GJ 9066c \citep{Feng2020,Quirrenbach2022} and GJ 3512b and c \citep{Morales2019,Morales2025}, which the standard core accretion mechanism faces challenges \citep{Liu2019,Burn2021}. These violent dynamical processes can be probed and examined by investigating the mutual inclinations of multiple planets. 


\section{Conclusions}\label{conclusion}

In this work, we jointly analyzed RVs and Hipparcos–Gaia astrometry for 19 M dwarfs hosting cold Jupiters found by long-term RV observations. 
We obtain robust absolute masses for 13 cold Jupiters in 11 systems, including the first such measurements for five planets. 
The combined data also show that HIP 79431B, previously identified as a planet based on its minimum mass, is a brown dwarf with a true mass of $31.3^{+8.5}_{-9.7}\,M_{\rm Jup}$. 
Several systems still have degenerate orbital solutions or poorly constrained inclinations, which Gaia DR4 epoch astrometry may help resolve. 
Comparing planet-to-star mass ratios of giant planets orbiting M stars found by different techniques, we find a possible paucity of high-mass-ratio hot Jupiters around M dwarfs, and our local cold-Jupiter sample resembles the microlensing cold-Jupiter sample, whose planets lie predominantly toward the Galactic bulge. 
These population comparisons remain tentative because of the limited sample size and uncorrected detection completeness. 
Our study provides a local sample of cold Jupiters around M dwarfs with well-constrained masses and orbits, and reveals a tip of a larger population of cold Jupiters expected from the upcoming Gaia DR4 and DR5.


\section{Acknowledgments}
We gratefully acknowledge the contributions of the PFS observers who participated in the observations used in this work, including Mauro Barbieri, Jennifer Burt, Yuri Beletsky, Karen Collins, Mat\'{i}as D\'{i}az, Diana Dragomir, Zahra Essack, Adina Feinstein, Erin Flowers, Shubham Kanodia, Adriana Kuehnel, Michelle Kunimoto, Ben Montet, Sam Quinn, Malena Rice, Marcelo Tala, Sam Yee, and George Zhou. Their efforts in obtaining these observations are greatly appreciated.

This work is supported by the National Key R\&D Program of China, No.2024YFA1611801 and No. 2024YFC2207700 by the National Natural Science Foundation of China (NSFC) under Grant No. 12473066 and 12573068, by the Shanghai Jiao Tong University 2030 Initiative, by SJTU-Warwick Joint Seed Fund 2024/25-Round 5. and by the China Chile Joint Research Fund (CCJRF No. 2205). CCJRF is provided by Chinese Academy of Sciences South America Center for Astronomy (CASSACA) and established by National Astronomical Observatories, Chinese Academy of Sciences (NAOC) and Chilean Astronomy Society (SOCHIAS) to support China-Chile collaborations in astronomy. The computations in this paper were run on the Siyuan-1 cluster supported by the Center for High Performance Computing at Shanghai Jiao Tong University. 

T.G. and E.P. acknowledges financial support from the Agencia Estatal de Investigaci\'on of the Ministerio de Ciencia e Innovaci\'on MCIN/AEI/10.13039/501100011033 and the ERDF “A way of making Europe” through projects PID2021-125627OB-C32 and PID2024-158486OB-C32. This work is supported by the ERC Grant (ERC Advanced Grant SPEAR, GA 101200674). 
P. A. acknowledges financial support from the Agencia Estatal de Investigación (AEI/10.13039/501100011033) of the Ministerio de Ciencia, Innovación y Universidades and the ERDF “A way of making Europe” through
Projects: (i) PID2022-137241NB-C43, with financing from the European Union – NextGenerationEU; the Ministry of Science, Innovation and Universities; the Recovery, Transformation and Resilience Plan; the Regional Ministry of University, Research and Innovation of the Regional Government of Andalusia; and the Spanish National Research Council, and (ii) the Centre of Excellence “Severo Ochoa” award to the Instituto de Astrofísica de Andalucía (CEX2021-001131-S).
J.C.M. and I. R. acknowledge support from Spanish grants PID2024-158486OB-C31 and PID2025-172309OB-I00 funded by MCIU/AEI/10.13039/501100011033 and by “ERDF A way of making Europe” and the European Social Fund Plus (ESF+), from the programme Unidad de Excelencia Mar\'ia de Maeztu CEX2020-001058-M financed by MCIN/AEI/10.13039/501100011033 and by the MaX-CSIC Excellence Award MaX4-SOMMA-ICE, from the Generalitat de Catalunya/CERCA programme, and from the European Research Council (ERC) under the European Union’s Horizon Europe programme (ERC Advanced Grant SPOTLESS; no. 101140786). Funded by the European Union. Views and opinions expressed are however those of the author(s) only and do not necessarily reflect those of the European Union or the  European Research Council Executive Agency (ERCEA). Neither the European Union nor the granting authority can be held responsible for them.

Based on observations collected at the Centro Astronómico Hispano en Andalucía
(CAHA) at Calar Alto, operated jointly by Junta de Andalucía and Consejo Superior
de Investigaciones Científicas (IAA-CSIC). This paper includes data gathered with the 6.5 meter Magellan Telescopes located at Las Campanas Observatory, Chile.

Generative AI tools were used to suggest language improvements to the manuscript and to assist with some of the data analysis and code development. All AI-assisted outputs were reviewed and verified by the authors, who retain full responsibility for the scientific content and conclusions of this work. The generative AI tool used was ChatGPT.
%

\vspace{5mm}
\facilities{Hipparcos, Gaia, CARMENES, HARPS, HARPS-N, HIRES, HRS, IRD, NIRPS, PFS, SOPHIE, Tull, UCLES}


\software{astropy \citep{exoplanet:astropy13,exoplanet:astropy18,exoplanet:astropy22}, corner \citep{Foreman2016},  matplotlib \citep{Hunter2007}, numpy \citep{numpy2011, numpy2020}, pandas \citep{pandas2022}, ptemcee \citep{Vousden2016}, scipy \citep{Virtanen2020}
          }




\appendix

\restartappendixnumbering

\section{Diagnostic plots of joint analysis}

Figures~\ref{fig:fit_GJ317}, \ref{fig:fit_GJ328}, \ref{fig:fit_GJ463}, \ref{fig:fit_GJ508.2}, \ref{fig:fit_GJ649}, \ref{fig:fit_GJ676A}, \ref{fig:fit_GJ832}, \ref{fig:fit_GJ849}, \ref{fig:fit_GJ9066}, \ref{fig:fit_HIP79431} and \ref{fig:fit_HIP88962} show the joint-fit results of GJ 317, GJ 328, GJ 463, GJ 508.2, GJ 649, GJ 676A, GJ 832, GJ 849, GJ 9066, HIP 79431 and HIP 88962.

\renewcommand{\theHfigure}{A.\arabic{figure}} 
\setcounter{figure}{0}

\begin{figure*}[htb]
    \centering
    \includegraphics[width=0.99\textwidth]{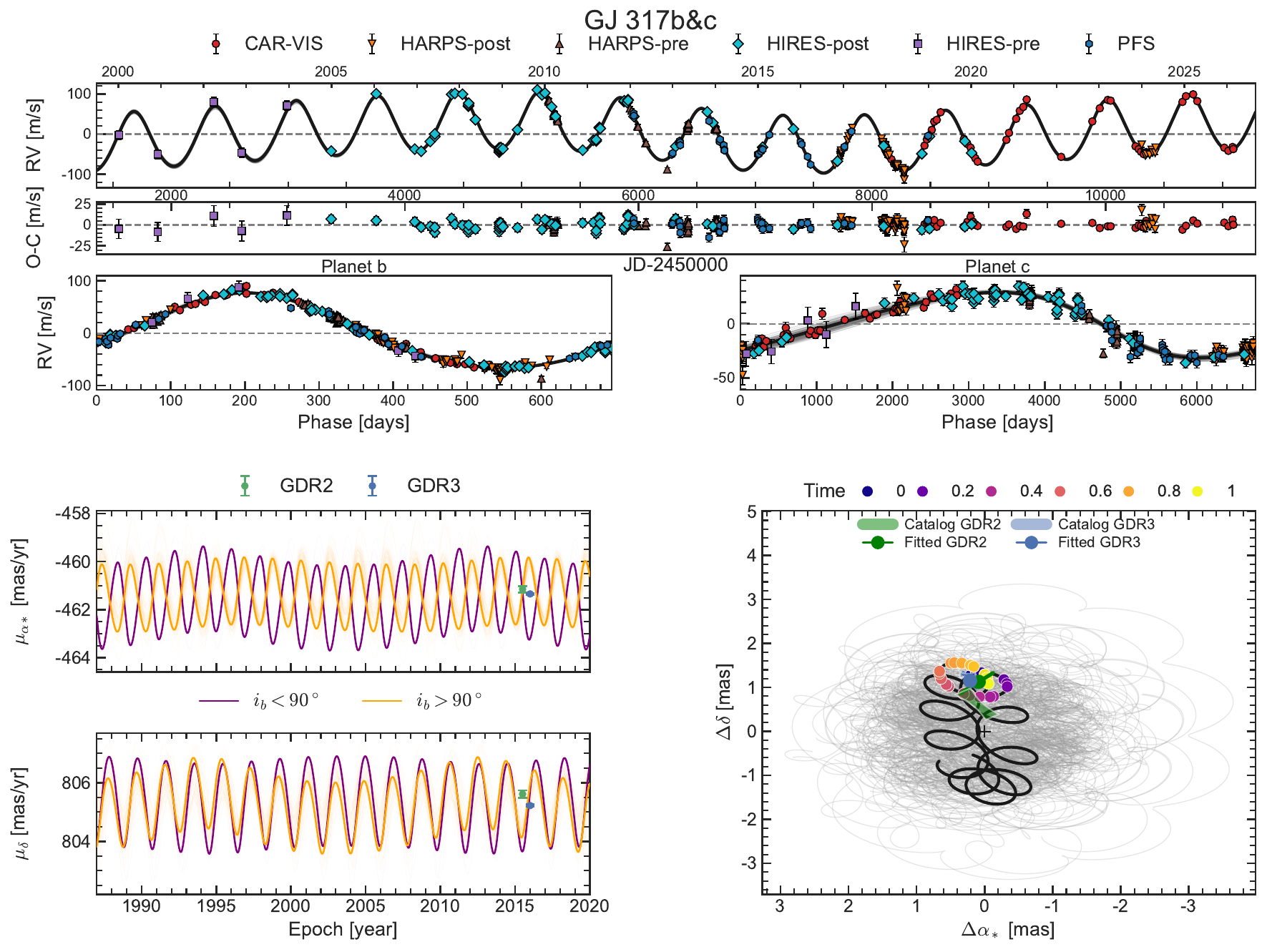}
    \caption{Same as Figure~\ref{fig:fit_GJ179} but for GJ 317.}
    \label{fig:fit_GJ317}
\end{figure*}

\begin{figure*}[htb]
    \centering
    \includegraphics[width=0.99\textwidth]{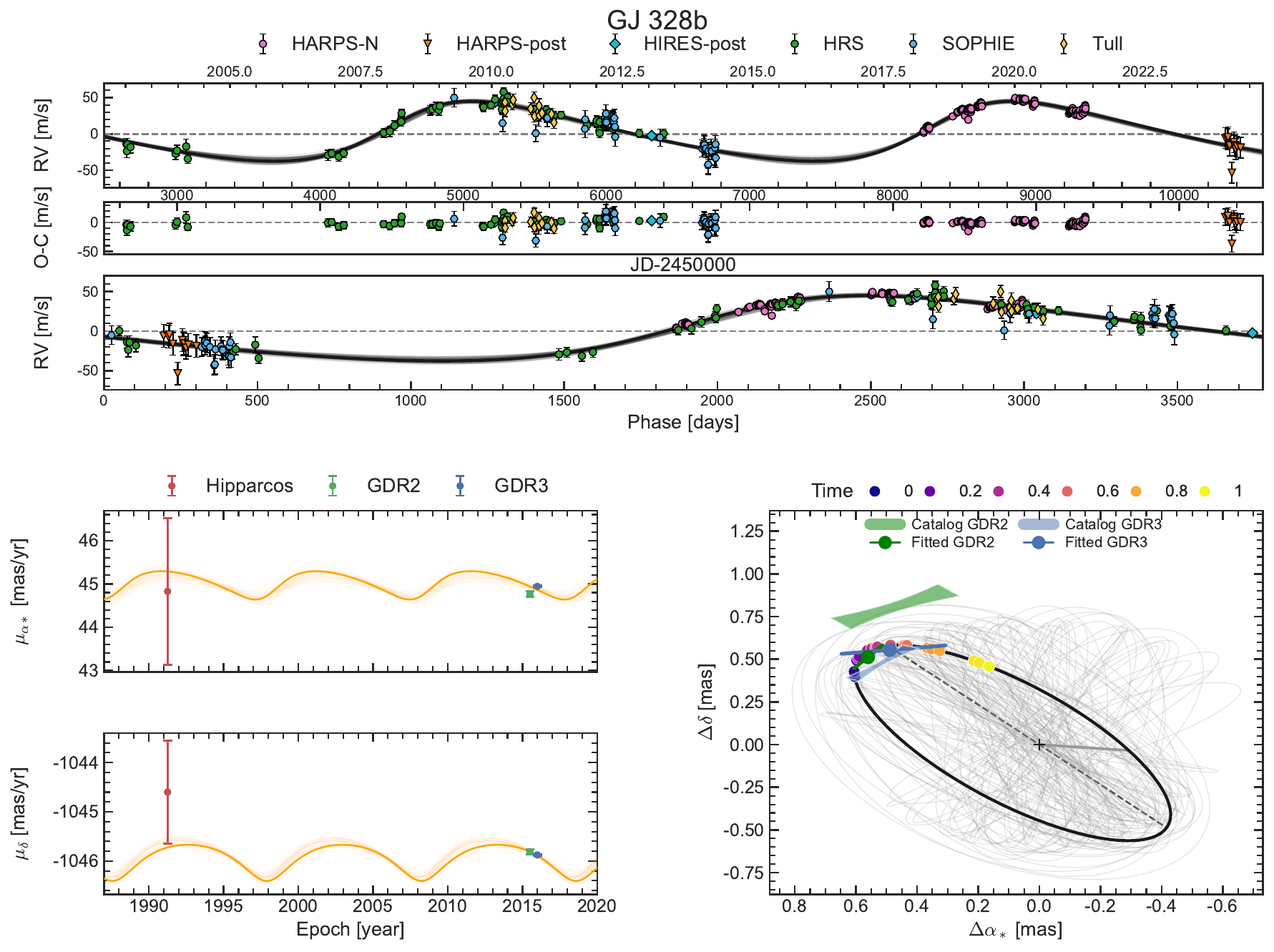}
    \caption{Same as Figure~\ref{fig:fit_GJ179} but for GJ 328.}
    \label{fig:fit_GJ328}
\end{figure*}

\begin{figure*}[htb]
    \centering
    \includegraphics[width=0.99\textwidth]{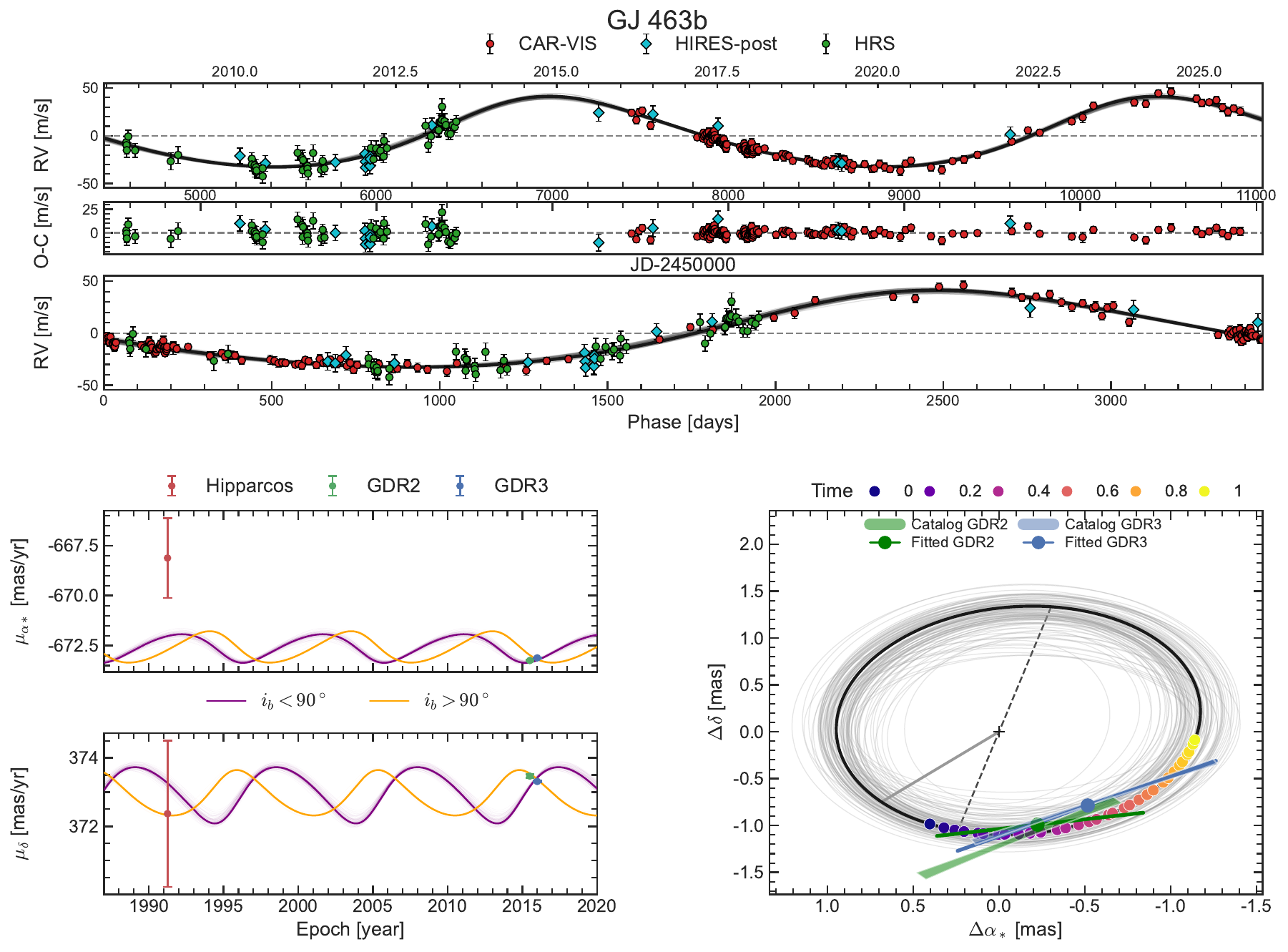}
    \caption{Same as Figure~\ref{fig:fit_GJ179} but for GJ 463.}
    \label{fig:fit_GJ463}
\end{figure*}

\begin{figure*}[htb]
    \centering
    \includegraphics[width=0.99\textwidth]{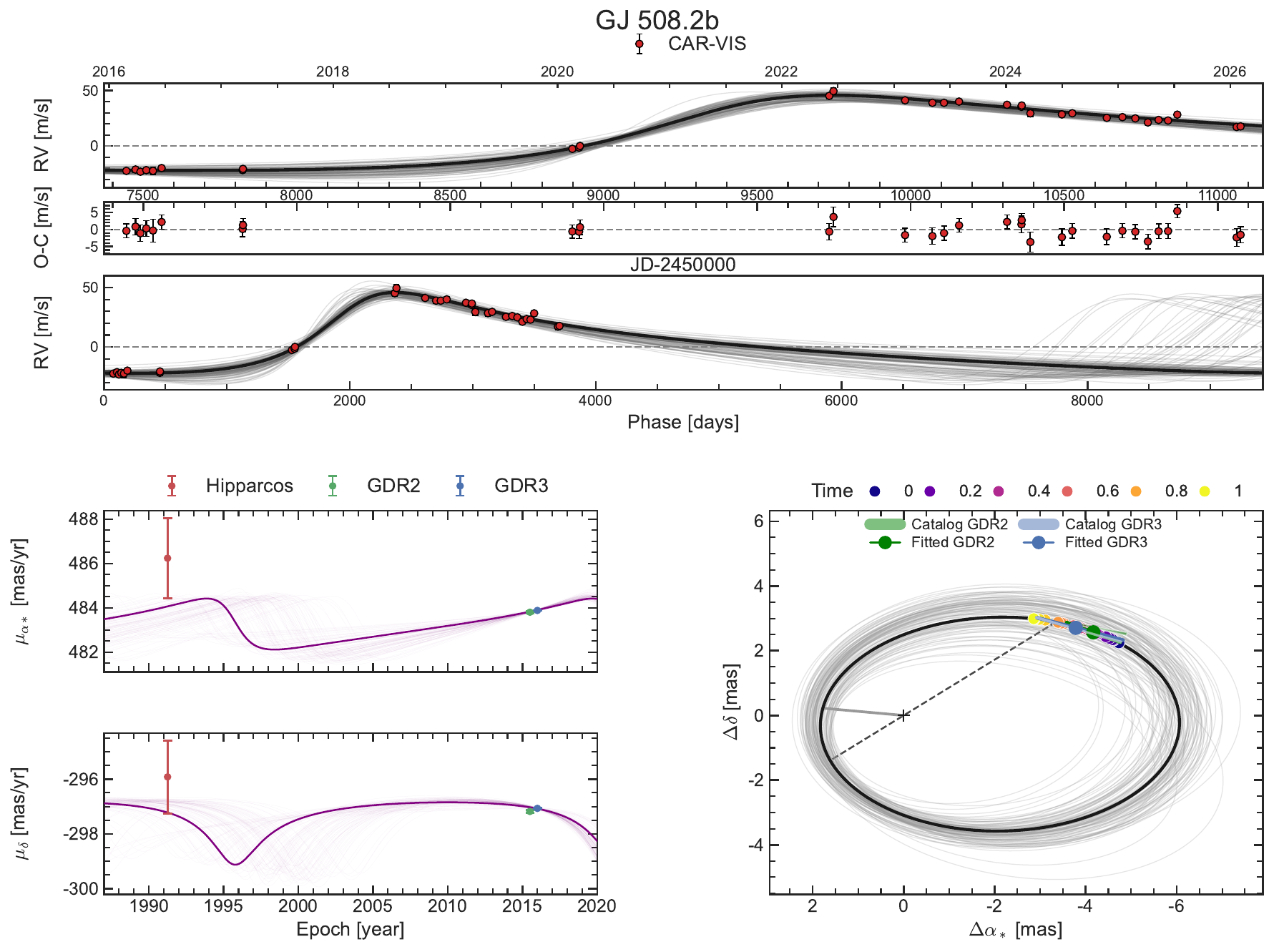}
    \caption{Same as Figure~\ref{fig:fit_GJ179} but for GJ 508.2.}
    \label{fig:fit_GJ508.2}
\end{figure*}

\begin{figure*}[htb]
    \centering
    \includegraphics[width=0.99\textwidth]{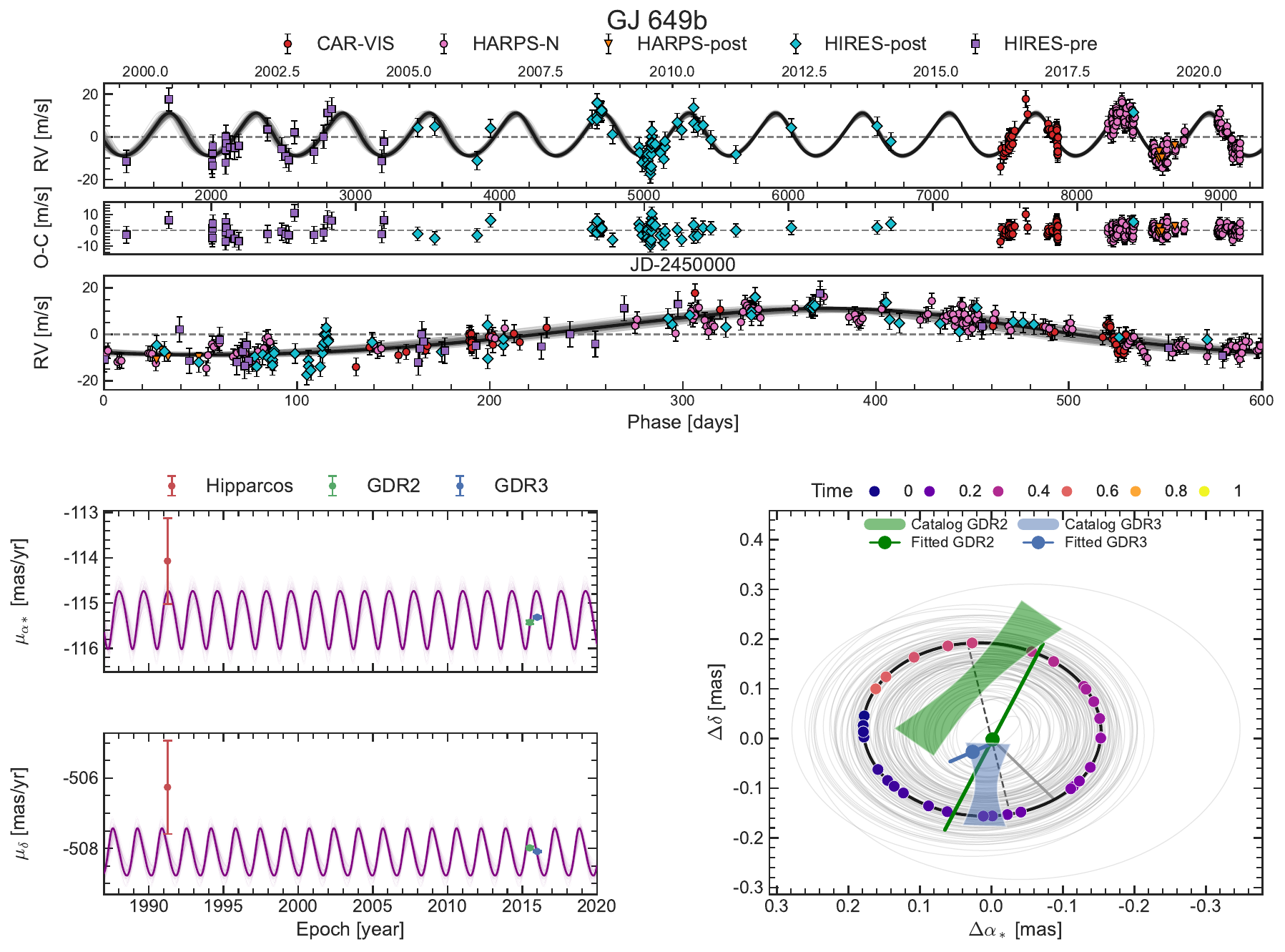}
    \caption{Same as Figure~\ref{fig:fit_GJ179} but for GJ 649.}
    \label{fig:fit_GJ649}
\end{figure*}

\begin{figure*}[htb]
    \centering
    \includegraphics[width=0.99\textwidth]{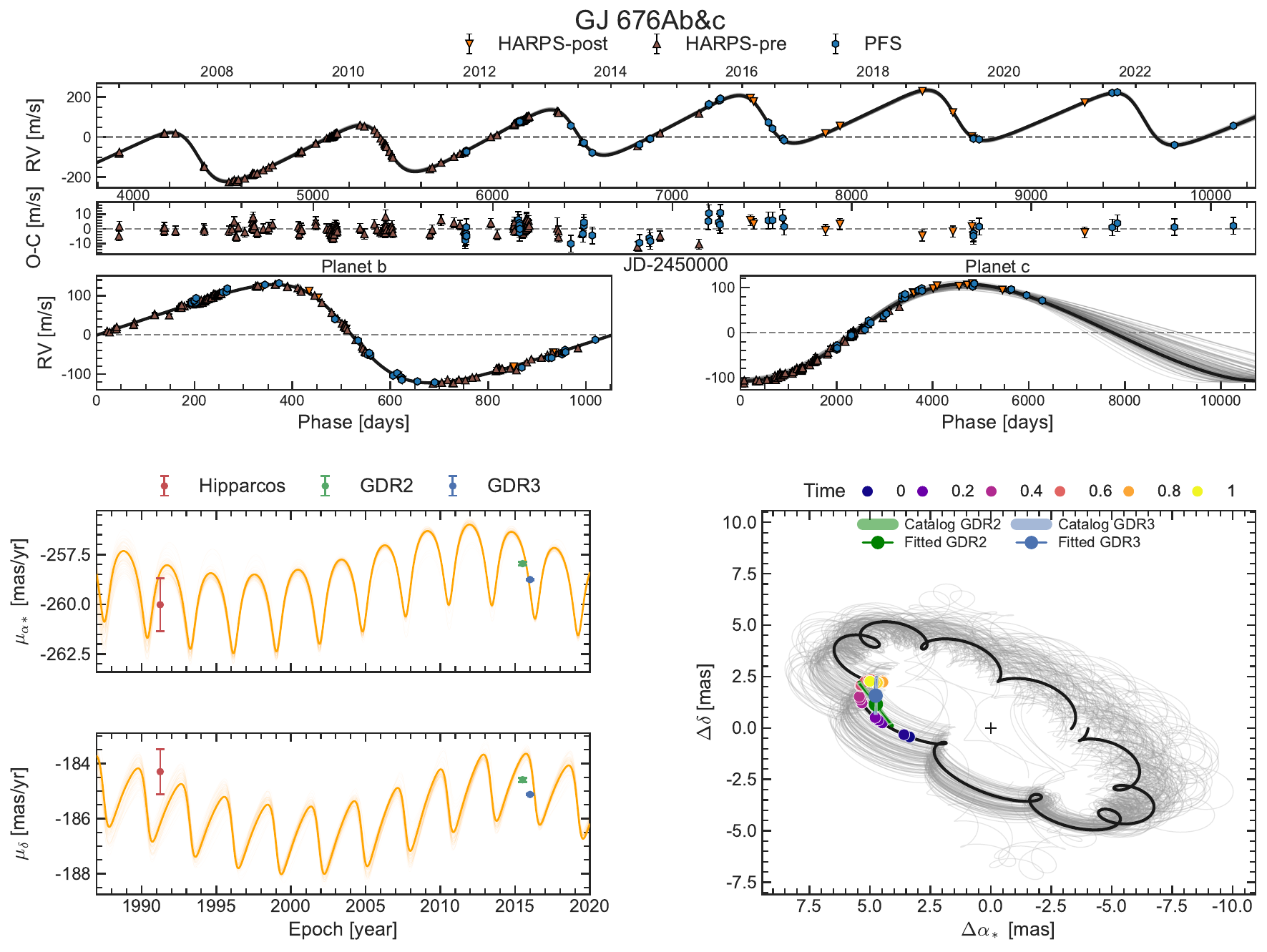}
    \caption{Same as Figure~\ref{fig:fit_GJ179} but for GJ 676A.}
    \label{fig:fit_GJ676A}
\end{figure*}

\begin{figure*}[htb]
    \centering
    \includegraphics[width=0.99\textwidth]{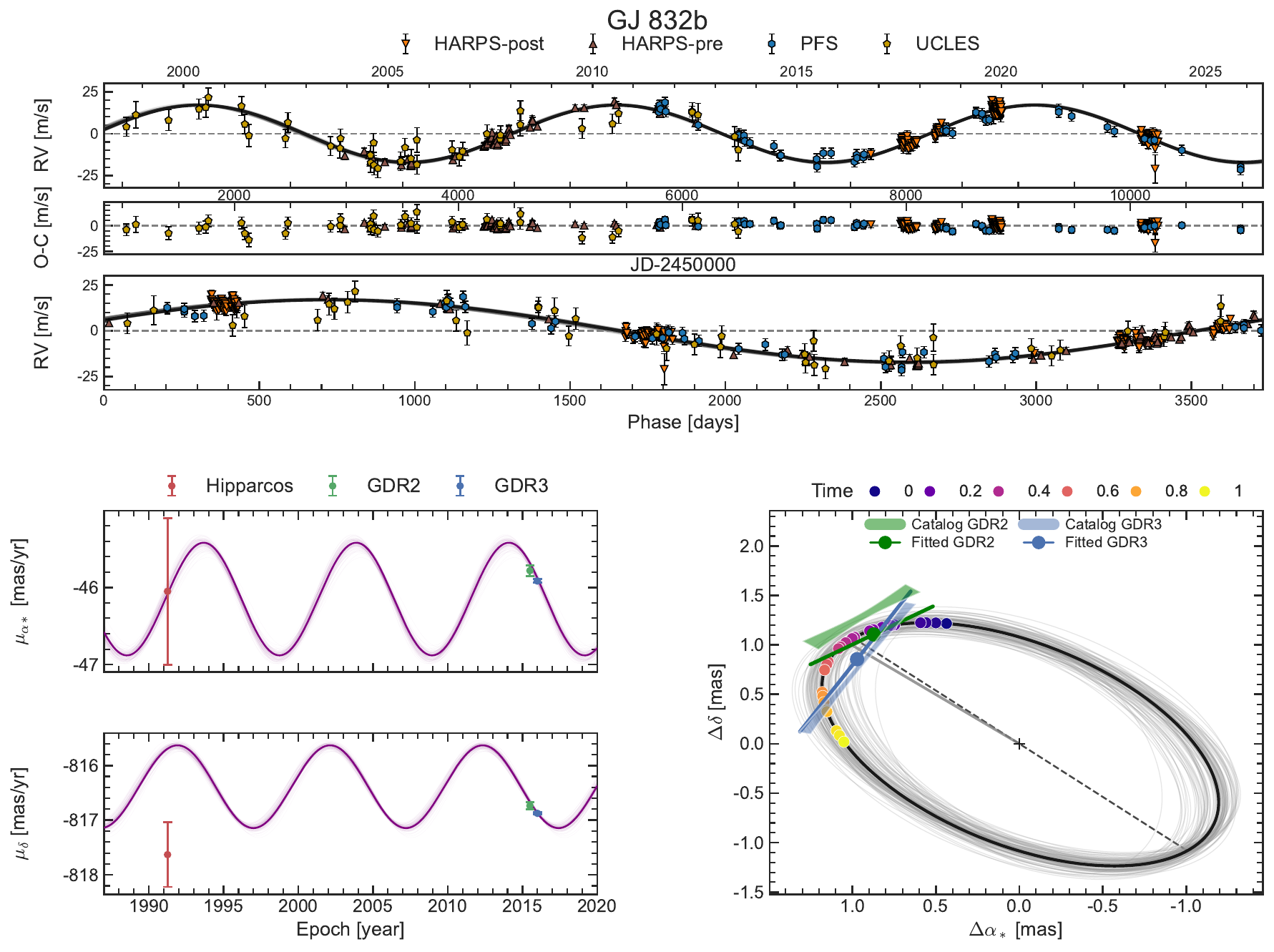}
    \caption{Same as Figure~\ref{fig:fit_GJ179} but for GJ 832.}
    \label{fig:fit_GJ832}
\end{figure*}

\begin{figure*}[htb]
    \centering
    \includegraphics[width=0.99\textwidth]{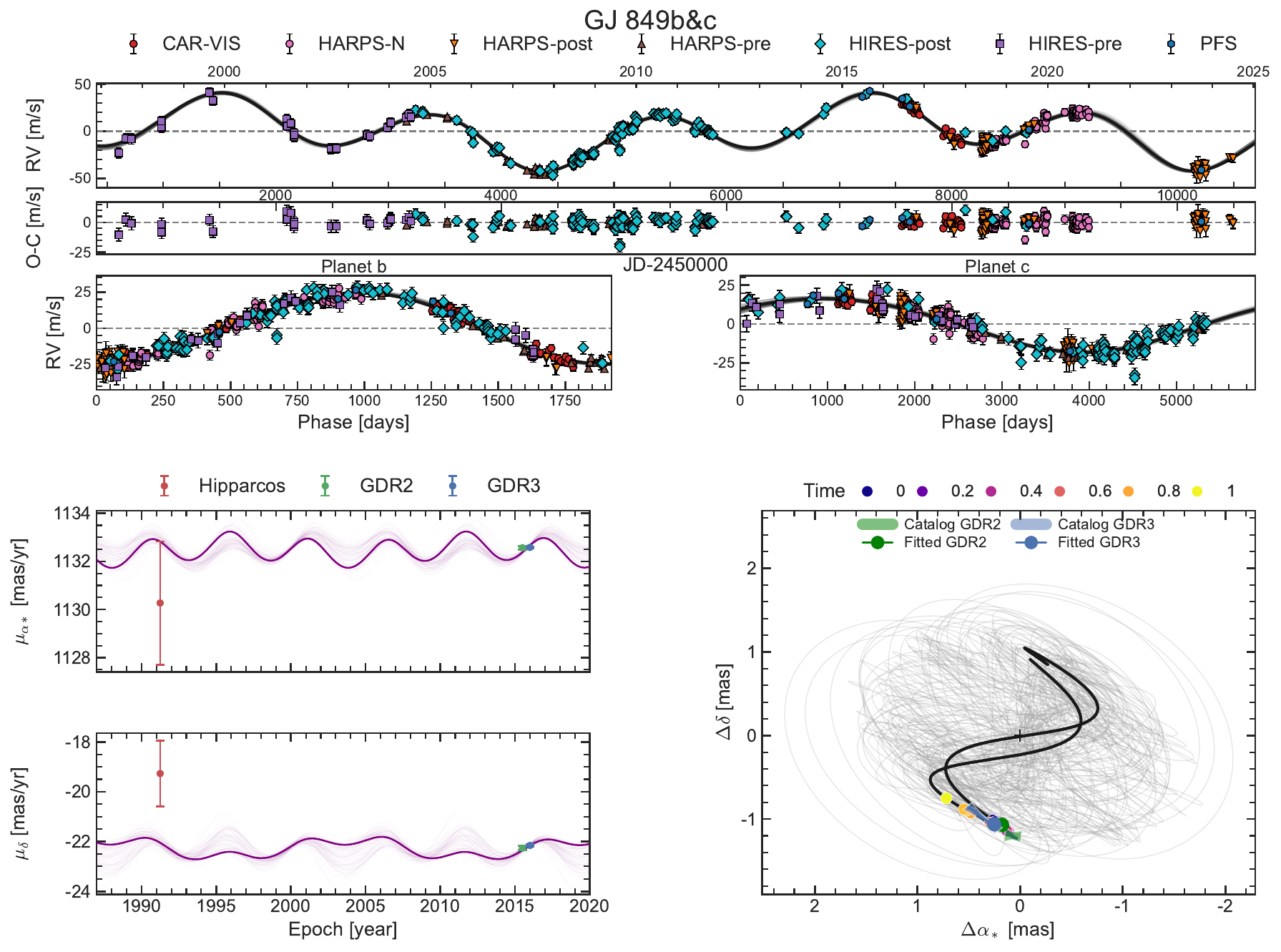}
    \caption{Same as Figure~\ref{fig:fit_GJ179} but for GJ 849.}
    \label{fig:fit_GJ849}
\end{figure*}

\begin{figure*}[htb]
    \centering
    \includegraphics[width=0.99\textwidth]{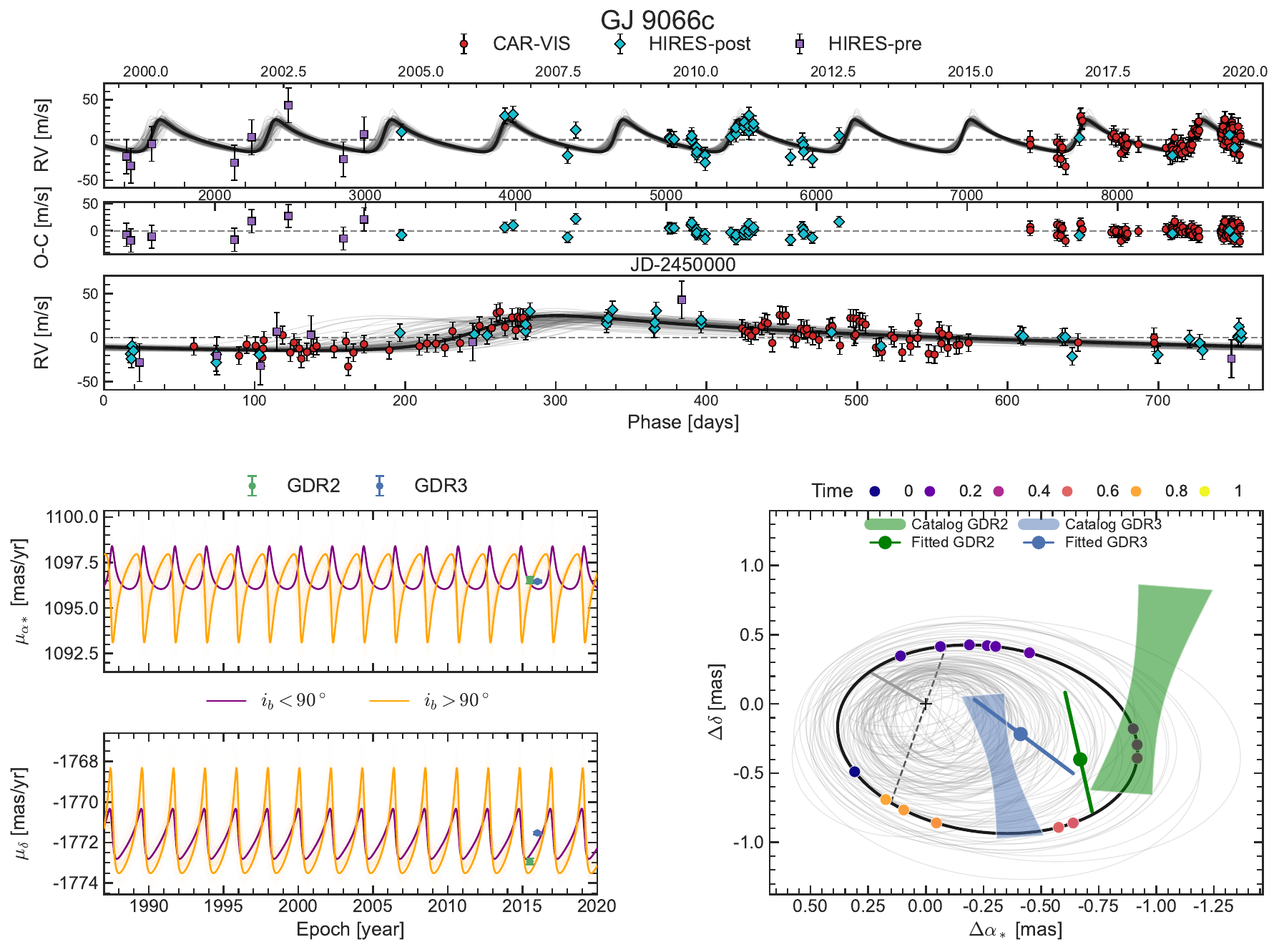}
    \caption{Same as Figure~\ref{fig:fit_GJ179} but for GJ 9066.}
    \label{fig:fit_GJ9066}
\end{figure*}

\begin{figure*}[htb]
    \centering
    \includegraphics[width=0.99\textwidth]{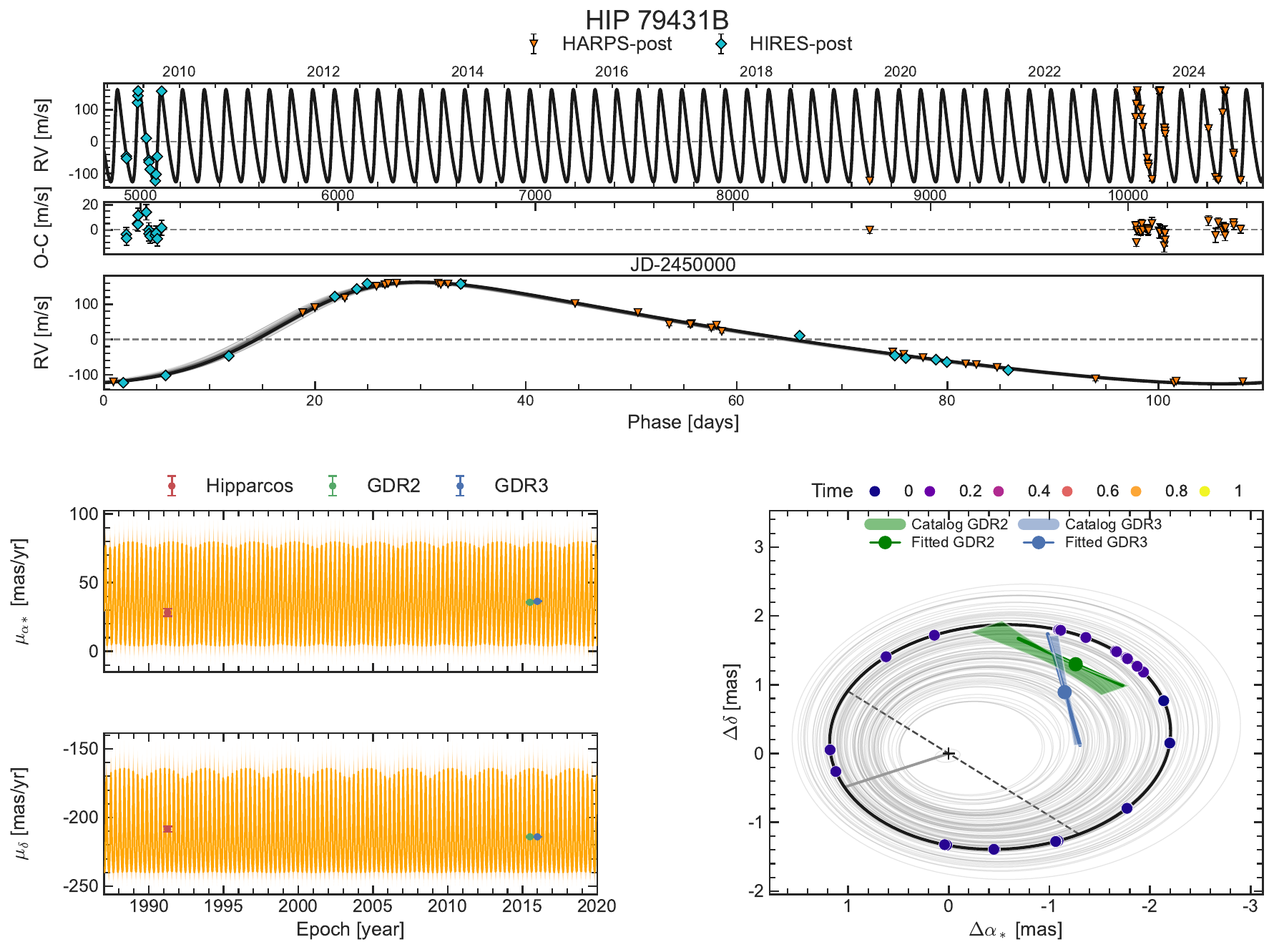}
    \caption{Same as Figure~\ref{fig:fit_GJ179} but for HIP 79431.}
    \label{fig:fit_HIP79431}
\end{figure*}

\begin{figure*}[htb]
    \centering
    \includegraphics[width=0.99\textwidth]{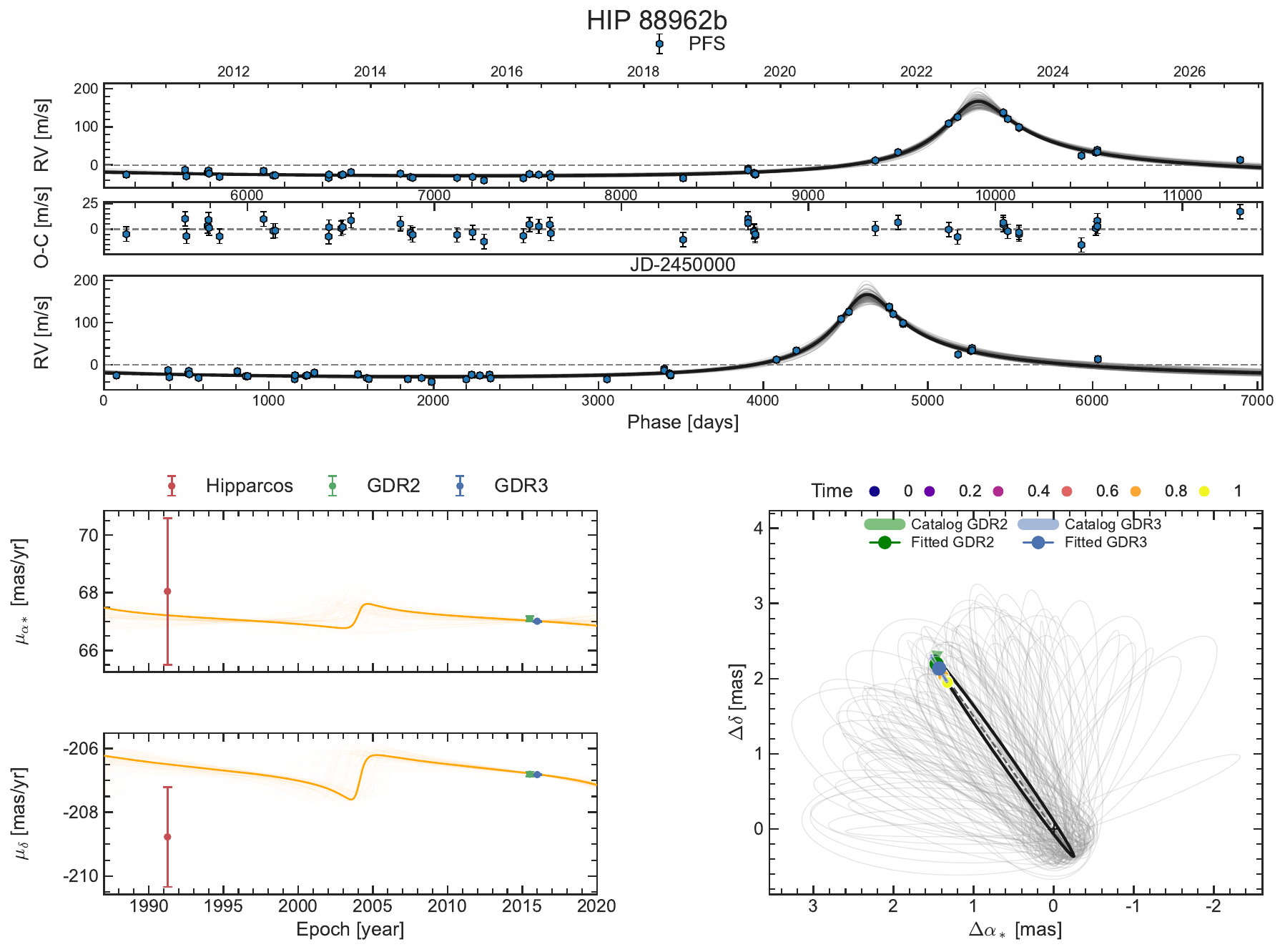}
    \caption{Same as Figure~\ref{fig:fit_GJ179} but for HIP 88962.}
    \label{fig:fit_HIP88962}
\end{figure*}

\section{Posterior plots of joint analysis}

Figures~\ref{fig:posterior_GJ179} show the posteriors of all 12 systems, including 11 planetary systems (GJ 179b, GJ 317b, GJ 328b, GJ 463b, GJ 508.2b, GJ 649b, GJ 676Ab and GJ 676Ac, GJ 832b, GJ 849b and GJ 849c, GJ 9066c and HIP 88962b) and one brown dwarf (HIP 79431B), that have robust orbital solutions. The complete figure set (12 plots) is available in the online journal.

\renewcommand{\theHfigure}{B.\arabic{figure}} 
\setcounter{figure}{0}

\figsetstart
\figsetnum{1}
\figsettitle{Posteriors of joint analysis}

\figsetgrpstart
\figsetgrpnum{1.1}
\figsetplot{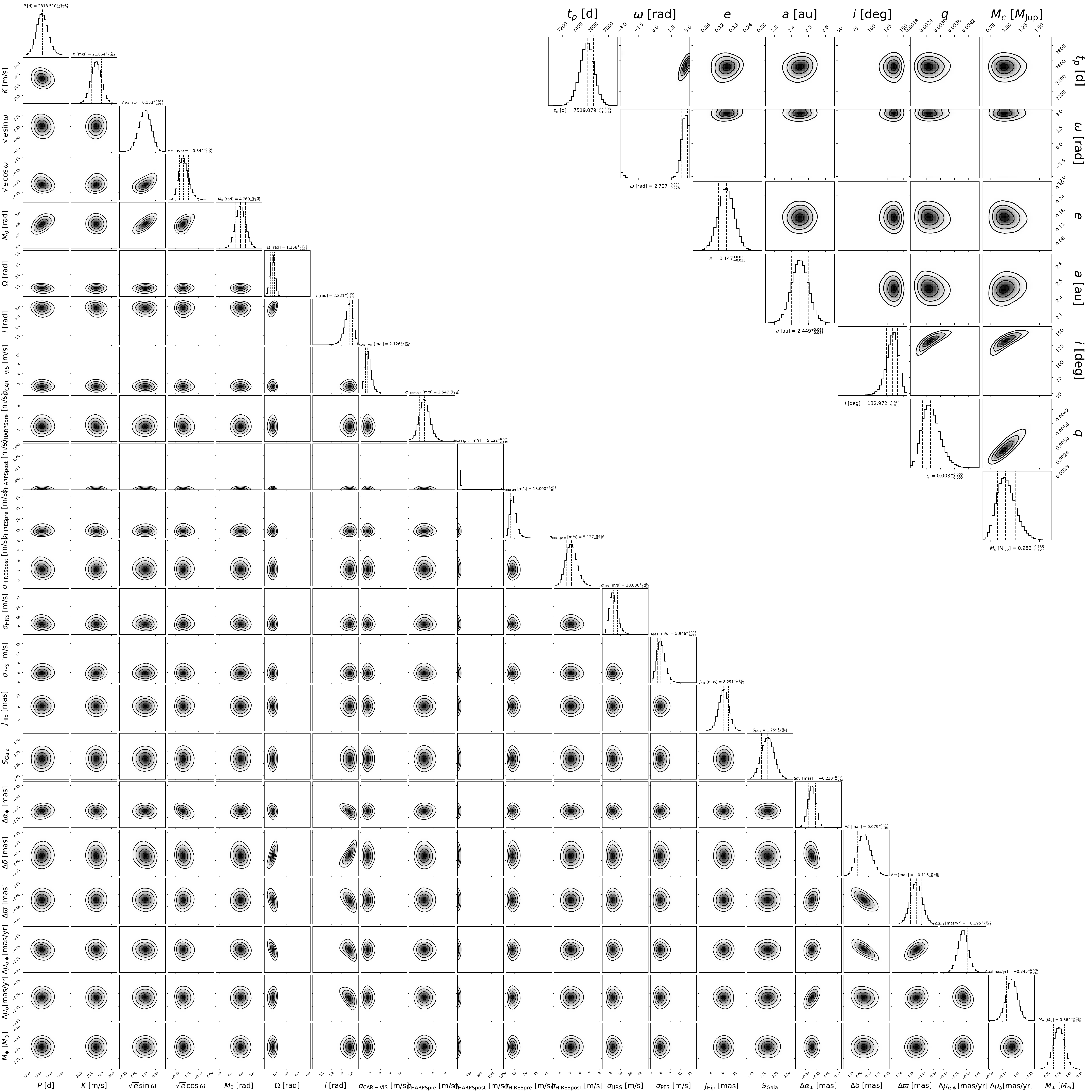}
\figsetgrpnote{Posterior distributions of the fitting parameters (lower left) and derived parameters (upper right) from RV + Astrometry joint fit for the GJ 179 system. The value of $t_p$ is relative to JD2,450,000.}
\figsetgrpend

\figsetend

\begin{figure*}
\epsscale{1.2}
\plotone{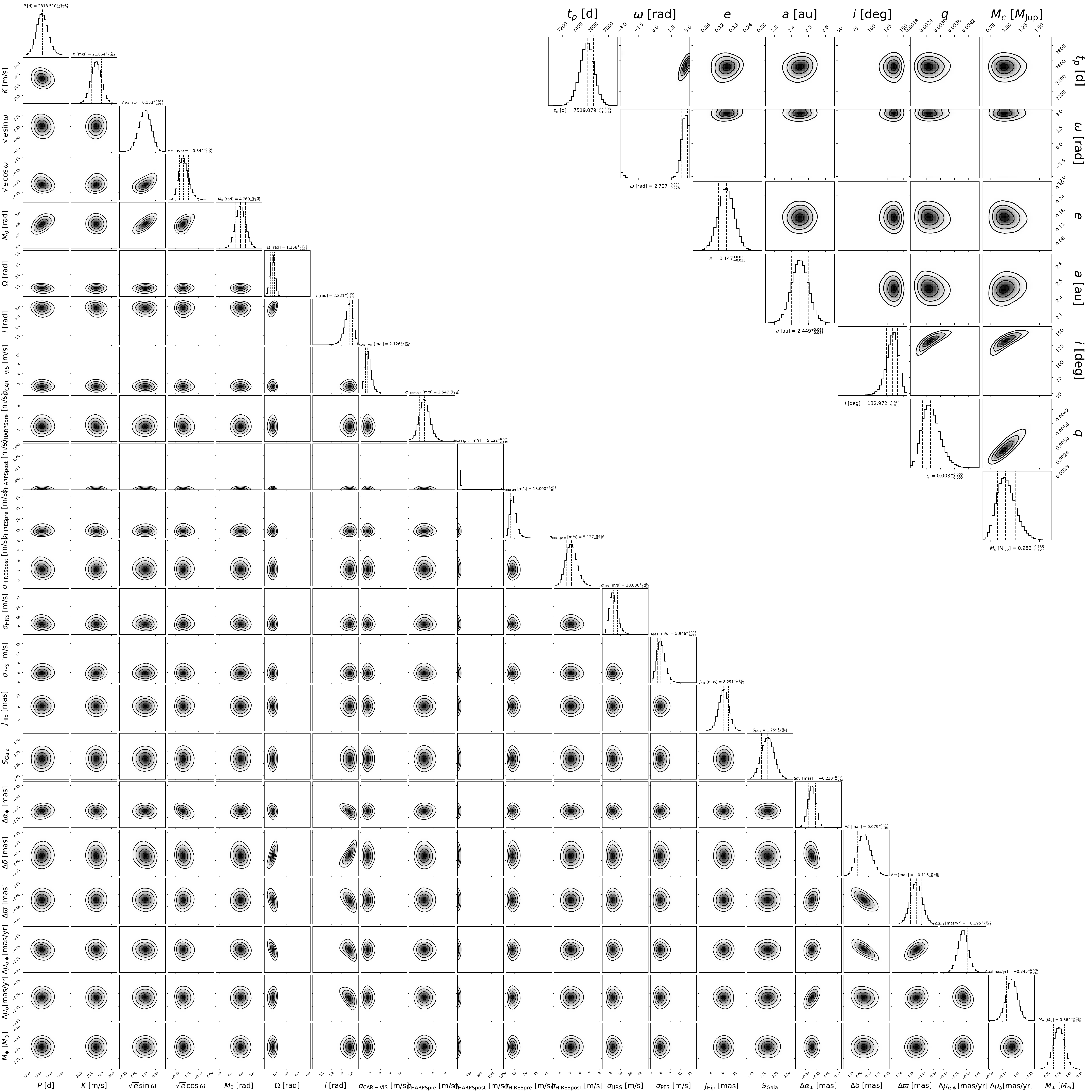}
\caption{Posterior distributions of the fitting parameters (lower left) and derived parameters (upper right) from RV + Astrometry joint fit for the GJ 179 system. The value of $t_p$ is relative to JD2,450,000. The complete figure set (12 plots) is available in the online journal.}
\label{fig:posterior_GJ179}
\end{figure*}

\section{Parameters of systems with robust orbital solutions}

Table~\ref{sample_result_summary} summarizes the best-fit and derived parameters of all systems that have robust orbital solutions with both upper and lower uncertainties on inclination smaller than 30 deg.

\renewcommand{\theHtable}{C.\arabic{table}}
\setcounter{table}{0}

\begin{table*}[htb]
\centering
{\renewcommand{\arraystretch}{1.0}
\setlength{\tabcolsep}{3pt}
\caption{Best-fit and derived parameters of all systems with robust orbital solutions}
\label{sample_result_summary}
\resizebox{0.97\textwidth}{!}{%
\begin{tabular}{rcccccccc}
\hline\hline
Parameter & GJ 179 & GJ 317 & GJ 317 & GJ 328 & GJ 463 & GJ 463 & GJ 508.2 & GJ 649 \\ 
  &  & $(i_b < 90^{\circ})$ & $(i_b > 90^{\circ})$ & & $(i_b < 90^{\circ})$ & $(i_b > 90^{\circ})$ & & \\ \hline
\multicolumn{9}{c}{Fitted parameters} \\ \hline
$P_{b} \ (\rm days)$ & $2318.5_{-23.9}^{+26.3}$ & $695.0_{-0.2}^{+0.5}$ & $695.0_{-0.3}^{+0.3}$ & $3766.8_{-17.7}^{+17.6}$ & $3464.6_{-20.9}^{+21.8}$ & $3463.6_{-19.5}^{+20.8}$ & $8591.8_{-1259.4}^{+1333.3}$ & $600.3_{-1.5}^{+1.5}$ \\
$K_{b} \ (\rm m\ s^{-1})$ & $21.9_{-0.7}^{+0.7}$ & $71.5_{-0.6}^{+0.5}$ & $71.6_{-0.6}^{+0.6}$ & $40.9_{-1.7}^{+1.7}$ & $36.6_{-0.7}^{+0.8}$ & $36.6_{-0.7}^{+0.7}$ & $34.3_{-1.0}^{+1.4}$ & $9.8_{-0.3}^{+0.3}$ \\
$\sqrt{e_{b}} \sin \omega_{b}$ & $0.15_{-0.09}^{+0.08}$ & $-0.01_{-0.02}^{+0.01}$ & $-0.00_{-0.02}^{+0.01}$ & $-0.42_{-0.03}^{+0.03}$ & $-0.20_{-0.04}^{+0.04}$ & $-0.20_{-0.04}^{+0.04}$ & $-0.57_{-0.05}^{+0.05}$ & $0.19_{-0.09}^{+0.07}$ \\
$\sqrt{e_{b}} \cos \omega_{b}$ & $-0.34_{-0.05}^{+0.07}$ & $0.28_{-0.02}^{+0.01}$ & $0.28_{-0.01}^{+0.01}$ & $0.21_{-0.05}^{+0.05}$ & $0.31_{-0.03}^{+0.02}$ & $0.31_{-0.03}^{+0.02}$ & $0.44_{-0.09}^{+0.07}$ & $0.27_{-0.11}^{+0.08}$ \\
$M_{0,b} \ (\rm rad)$ & $1.13_{-0.24}^{+0.26}$ & $0.06_{-0.05}^{+0.06}$ & $0.04_{-0.03}^{+0.06}$ & $5.52_{-0.09}^{+0.09}$ & $0.35_{-0.15}^{+0.16}$ & $0.35_{-0.15}^{+0.16}$ & $5.25_{-0.15}^{+0.12}$ & $1.83_{-0.36}^{+0.28}$ \\
$\Omega_{b} \ (\rm rad)$ & $1.16_{-0.28}^{+0.24}$ & $1.82_{-0.20}^{+0.14}$ & $4.38_{-0.34}^{+0.39}$ & $3.63_{-0.67}^{+0.34}$ & $2.90_{-0.09}^{+0.09}$ & $1.50_{-0.10}^{+0.11}$ & $2.39_{-0.21}^{+0.17}$ & $3.21_{-0.32}^{+0.25}$ \\
$i_{b} \ (\rm rad)$ & $2.32_{-0.17}^{+0.14}$ & $0.66_{-0.12}^{+0.18}$ & $2.28_{-0.18}^{+0.13}$ & $1.69_{-0.27}^{+0.23}$ & $0.54_{-0.05}^{+0.06}$ & $2.54_{-0.07}^{+0.05}$ & $0.34_{-0.04}^{+0.06}$ & $0.32_{-0.07}^{+0.13}$ \\
$P_{c} \ (\rm days)$ & -- & $6656.7_{-170.4}^{+181.0}$ & $6777.2_{-171.2}^{+182.8}$ & -- & -- & -- & -- & -- \\
$K_{c} \ (\rm m\ s^{-1})$ & -- & $30.1_{-0.7}^{+0.8}$ & $30.3_{-0.7}^{+0.7}$ & -- & -- & -- & -- & -- \\
$\sqrt{e_{c}} \sin \omega_{c}$ & -- & $0.40_{-0.02}^{+0.02}$ & $0.41_{-0.02}^{+0.02}$ & -- & -- & -- & -- & -- \\
$\sqrt{e_{c}} \cos \omega_{c}$ & -- & $-0.07_{-0.05}^{+0.08}$ & $-0.08_{-0.05}^{+0.06}$ & -- & -- & -- & -- & -- \\
$M_{0,c} \ (\rm rad)$ & -- & $1.49_{-0.15}^{+0.17}$ & $1.44_{-0.13}^{+0.14}$ & -- & -- & -- & -- & -- \\
$\Omega_{c} \ (\rm rad)$ & -- & $3.97_{-3.45}^{+1.27}$ & $2.44_{-1.66}^{+2.82}$ & -- & -- & -- & -- & -- \\
$i_{c} \ (\rm rad)$ & -- & $1.44_{-0.66}^{+0.82}$ & $1.60_{-0.63}^{+0.63}$ & -- & -- & -- & -- & -- \\
$\sigma_{\rm CAR-VIS} \ (\rm m\ s^{-1})$ & $2.13_{-0.69}^{+0.95}$ & $2.85_{-0.50}^{+0.26}$ & $2.51_{-0.42}^{+0.50}$ & -- & $0.37_{-0.26}^{+0.40}$ & $0.39_{-0.28}^{+0.38}$ & $1.67_{-0.37}^{+0.42}$ & $2.78_{-0.35}^{+0.41}$ \\
$\sigma_{\rm HARPS-N} \ (\rm m\ s^{-1})$ & -- & -- & -- & $3.15_{-0.21}^{+0.24}$ & -- & -- & -- & $3.10_{-0.19}^{+0.20}$ \\
$\sigma_{\rm HARPS-pre} \ (\rm m\ s^{-1})$ & $2.55_{-0.80}^{+0.89}$ & $4.18_{-0.26}^{+0.55}$ & $4.21_{-0.39}^{+0.42}$ & -- & -- & -- & -- & -- \\
$\sigma_{\rm HARPS-post} \ (\rm m\ s^{-1})$ & $5.12_{-3.70}^{+8.44}$ & $0.82_{-0.43}^{+0.88}$ & $0.81_{-0.56}^{+0.80}$ & $14.11_{-2.86}^{+4.29}$ & -- & -- & -- & $2.47_{-0.90}^{+1.82}$ \\
$\sigma_{\rm HIRES-pre} \ (\rm m\ s^{-1})$ & $13.00_{-3.07}^{+4.44}$ & $10.82_{-4.38}^{+8.98}$ & $11.15_{-4.74}^{+7.99}$ & -- & -- & -- & -- & $4.94_{-0.75}^{+0.97}$ \\
$\sigma_{\rm HIRES-post} \ (\rm m\ s^{-1})$ & $5.13_{-0.52}^{+0.58}$ & $4.91_{-0.57}^{+0.78}$ & $4.91_{-0.58}^{+0.68}$ & $3.87_{-2.70}^{+6.15}$ & $7.97_{-1.69}^{+2.19}$ & $7.86_{-1.68}^{+2.25}$ & -- & $4.29_{-0.41}^{+0.48}$ \\
$\sigma_{\rm HRS} \ (\rm m\ s^{-1})$ & $10.04_{-2.63}^{+3.50}$ & -- & -- & $1.84_{-1.26}^{+1.48}$ & $1.50_{-1.04}^{+1.45}$ & $1.41_{-0.98}^{+1.45}$ & -- & -- \\
$\sigma_{\rm PFS} \ (\rm m\ s^{-1})$ & $5.95_{-1.06}^{+1.36}$ & $4.63_{-0.65}^{+0.64}$ & $4.56_{-0.82}^{+0.97}$ & -- & -- & -- & -- & -- \\
$\sigma_{\rm SOPHIE} \ (\rm m\ s^{-1})$ & -- & -- & -- & $11.53_{-1.43}^{+1.73}$ & -- & -- & -- & -- \\
$\sigma_{\rm Tull} \ (\rm m\ s^{-1})$ & -- & -- & -- & $5.29_{-2.38}^{+2.57}$ & -- & -- & -- & -- \\
$\sigma_{\rm UCLES} \ (\rm m\ s^{-1})$ & -- & -- & -- & -- & -- & -- & -- & -- \\
$J_{\rm hip} \ (\rm mas)$ & $8.29_{-1.60}^{+1.59}$ & -- & -- & $1.20_{-0.84}^{+1.21}$ & $6.09_{-2.04}^{+1.78}$ & $6.30_{-2.05}^{+1.65}$ & $4.92_{-1.59}^{+1.33}$ & $5.27_{-0.78}^{+0.81}$ \\
$S_{\rm gaia}$ & $1.26_{-0.08}^{+0.08}$ & $1.39_{-0.09}^{+0.06}$ & $1.34_{-0.07}^{+0.08}$ & $1.92_{-0.06}^{+0.05}$ & $1.49_{-0.07}^{+0.07}$ & $1.50_{-0.07}^{+0.07}$ & $1.26_{-0.08}^{+0.08}$ & $1.21_{-0.08}^{+0.08}$ \\
$\Delta \alpha* \ (\rm mas)$ & $-0.21_{-0.05}^{+0.05}$ & $0.76_{-1.71}^{+0.61}$ & $-0.34_{-0.69}^{+1.20}$ & $0.36_{-0.46}^{+0.18}$ & $-0.53_{-0.10}^{+0.10}$ & $0.52_{-0.08}^{+0.08}$ & $-3.88_{-0.35}^{+0.34}$ & $0.01_{-0.02}^{+0.02}$ \\
$\Delta \delta \ (\rm mas)$ & $0.08_{-0.10}^{+0.11}$ & $-0.34_{-0.67}^{+1.39}$ & $0.03_{-0.95}^{+1.21}$ & $0.51_{-0.14}^{+0.08}$ & $-0.74_{-0.09}^{+0.09}$ & $0.70_{-0.10}^{+0.10}$ & $2.79_{-0.57}^{+0.44}$ & $-0.10_{-0.02}^{+0.02}$ \\
$\Delta \varpi \ (\rm mas)$ & $-0.12_{-0.04}^{+0.04}$ & $-0.00_{-0.07}^{+0.04}$ & $0.10_{-0.05}^{+0.05}$ & $0.01_{-0.03}^{+0.03}$ & $-0.01_{-0.03}^{+0.03}$ & $-0.05_{-0.02}^{+0.02}$ & $0.03_{-0.02}^{+0.02}$ & $-0.04_{-0.02}^{+0.02}$ \\
$\Delta \mu_{\alpha*} \ (\rm mas \ yr^{-1})$ & $-0.20_{-0.07}^{+0.06}$ & $0.09_{-0.25}^{+0.20}$ & $-0.10_{-0.21}^{+0.24}$ & $-0.02_{-0.07}^{+0.10}$ & $-0.54_{-0.06}^{+0.06}$ & $-0.47_{-0.06}^{+0.06}$ & $0.78_{-0.12}^{+0.09}$ & $-0.00_{-0.02}^{+0.02}$ \\
$\Delta \mu_{\delta} \ (\rm mas \ yr^{-1})$ & $-0.34_{-0.06}^{+0.06}$ & $-0.02_{-0.21}^{+0.38}$ & $-0.10_{-0.19}^{+0.24}$ & $0.01_{-0.04}^{+0.05}$ & $0.28_{-0.06}^{+0.06}$ & $0.38_{-0.07}^{+0.06}$ & $0.36_{-0.06}^{+0.10}$ & $0.06_{-0.03}^{+0.03}$ \\ \hline
\multicolumn{9}{c}{Derived parameters} \\ \hline
$i_{b} \ (\rm deg)$ & $133.0_{-9.9}^{+7.8}$ & $37.8_{-6.9}^{+10.3}$ & $130.7_{-10.1}^{+7.4}$ & $96.9_{-15.3}^{+13.0}$ & $31.1_{-2.7}^{+3.2}$ & $145.8_{-3.7}^{+3.0}$ & $19.2_{-2.0}^{+3.4}$ & $18.4_{-4.0}^{+7.4}$ \\
$m_{b} \ (M_{\rm Jup})$ & $0.98_{-0.13}^{+0.16}$ & $2.85_{-0.48}^{+0.51}$ & $2.32_{-0.28}^{+0.33}$ & $2.31_{-0.13}^{+0.15}$ & $3.18_{-0.27}^{+0.28}$ & $2.92_{-0.26}^{+0.25}$ & $5.83_{-1.09}^{+1.13}$ & $0.83_{-0.21}^{+0.22}$ \\
$q_{b}$ & $0.0026_{-0.0003}^{+0.0004}$ & $0.0065_{-0.0011}^{+0.0014}$ & $0.0052_{-0.0006}^{+0.0007}$ & $0.0035_{-0.0002}^{+0.0002}$ & $0.0064_{-0.0005}^{+0.0005}$ & $0.0059_{-0.0005}^{+0.0005}$ & $0.0105_{-0.0015}^{+0.0015}$ & $0.0016_{-0.0004}^{+0.0004}$ \\
$a_{b} \ (\rm AU)$ & $2.45_{-0.05}^{+0.05}$ & $1.16_{-0.03}^{+0.01}$ & $1.15_{-0.02}^{+0.02}$ & $4.06_{-0.04}^{+0.04}$ & $3.50_{-0.05}^{+0.05}$ & $3.50_{-0.05}^{+0.05}$ & $6.66_{-0.82}^{+0.87}$ & $1.12_{-0.01}^{+0.01}$ \\
$e_{b}$ & $0.147_{-0.033}^{+0.033}$ & $0.078_{-0.009}^{+0.007}$ & $0.079_{-0.008}^{+0.008}$ & $0.228_{-0.015}^{+0.015}$ & $0.138_{-0.016}^{+0.016}$ & $0.138_{-0.016}^{+0.016}$ & $0.530_{-0.059}^{+0.050}$ & $0.118_{-0.040}^{+0.040}$ \\
$\omega_{b} \ (\rm rad)$ & $2.71_{-0.28}^{+0.22}$ & $-0.03_{-0.06}^{+0.04}$ & $-0.01_{-0.06}^{+0.03}$ & $-1.10_{-0.12}^{+0.11}$ & $-0.58_{-0.10}^{+0.11}$ & $-0.57_{-0.11}^{+0.11}$ & $-0.92_{-0.13}^{+0.11}$ & $0.61_{-0.29}^{+0.39}$ \\
$T_{p,b} - 2450000 \ (\rm JD)$ & $7519.1_{-92.3}^{+85.9}$ & $7930.4_{-7.0}^{+5.2}$ & $7932.3_{-6.7}^{+3.4}$ & $4627.9_{-60.8}^{+57.7}$ & $6737.9_{-62.0}^{+63.8}$ & $6740.9_{-64.3}^{+65.2}$ & $762.8_{-1301.1}^{+1213.4}$ & $7762.2_{-27.1}^{+35.0}$ \\
$i_{c} \ (\rm deg)$ & -- & $82.2_{-38.0}^{+46.7}$$^{[1]}$ & $91.7_{-36.0}^{+36.0}$$^{[1]}$ & -- & -- & -- & -- & -- \\
$m_{c} \ (M_{\rm Jup})$ & -- & $1.91_{-0.31}^{+0.45}$ & $1.75_{-0.17}^{+0.55}$ & -- & -- & -- & -- & -- \\
$q_{c}$ & -- & $0.0044_{-0.0008}^{+0.0011}$ & $0.0040_{-0.0004}^{+0.0012}$ & -- & -- & -- & -- & -- \\
$a_{c} \ (\rm AU)$ & -- & $5.22_{-0.19}^{+0.14}$ & $5.28_{-0.12}^{+0.12}$ & -- & -- & -- & -- & -- \\
$e_{c}$ & -- & $0.170_{-0.016}^{+0.016}$ & $0.176_{-0.015}^{+0.016}$ & -- & -- & -- & -- & -- \\
$\omega_{c} \ (\rm rad)$ & -- & $1.74_{-0.19}^{+0.13}$ & $1.76_{-0.13}^{+0.13}$ & -- & -- & -- & -- & -- \\
$T_{p,c} - 2450000 \ (\rm JD)$ & -- & $6361.9_{-157.0}^{+124.3}$ & $6388.5_{-116.7}^{+118.3}$ & -- & -- & -- & -- & -- \\
\end{tabular}}}
\begin{tablenotes}
    \item[1] [1] The lower and upper uncertainties on the inclination of GJ 317c are both higher than $30^\circ$ so its orbital solution is not robust according to our definition in Section~\ref{results}. 
\end{tablenotes}
\end{table*}

\begin{table*}[htb]
\centering
{\renewcommand{\arraystretch}{1.0}
\setlength{\tabcolsep}{3pt}
\noindent \textbf{Table C1.} (continued)\par
\resizebox{0.97\textwidth}{!}{
\begin{tabular}{rccccccc}
\hline\hline
Parameter & GJ 676A & GJ 832 & GJ 849 & GJ 9066 & GJ 9066 & HIP 79431 & HIP 88962 \\ 
  &  &  &  & $(i_c < 90^{\circ})$ & $(i_c > 90^{\circ})$ & & \\ \hline
\multicolumn{8}{c}{Fitted parameters} \\ \hline
$P_{b} \ (\rm days)$ & $1052.9_{-0.6}^{+0.6}$ & $3736.7_{-13.2}^{+13.1}$ & $1921.6_{-4.3}^{+4.3}$ & -- & -- & $109.919_{-0.008}^{+0.008}$ & $7054.9_{-482.3}^{+693.3}$ \\
$K_{b} \ (\rm m\ s^{-1})$ & $125.4_{-0.6}^{+0.6}$ & $17.2_{-0.3}^{+0.3}$ & $24.8_{-0.4}^{+0.3}$ & -- & -- & $143.8_{-1.3}^{+1.4}$ & $94.9_{-4.3}^{+5.2}$ \\
$\sqrt{e_{b}} \sin \omega_{b}$ & $0.561_{-0.004}^{+0.003}$ & $-0.01_{-0.09}^{+0.09}$ & $0.21_{-0.03}^{+0.03}$ & -- & -- & $-0.53_{-0.01}^{+0.01}$ & $-0.17_{-0.03}^{+0.03}$ \\
$\sqrt{e_{b}} \cos \omega_{b}$ & $0.04_{-0.01}^{+0.01}$ & $-0.06_{-0.06}^{+0.11}$ & $-0.05_{-0.04}^{+0.04}$ & -- & -- & $0.22_{-0.01}^{+0.01}$ & $0.83_{-0.02}^{+0.02}$ \\
$M_{0,b} \ (\rm rad)$ & $2.53_{-0.01}^{+0.01}$ & $1.28_{-0.95}^{+2.35}$ & $0.38_{-0.21}^{+0.22}$ & -- & -- & $5.56_{-0.02}^{+0.02}$ & $4.55_{-0.13}^{+0.16}$ \\
$\Omega_{b} \ (\rm rad)$ & $1.33_{-0.06}^{+0.07}$ & $3.89_{-0.09}^{+0.09}$ & $4.09_{-0.34}^{+0.33}$ & -- & -- & $0.82_{-0.09}^{+0.07}$ & $3.67_{-0.59}^{+0.63}$ \\
$i_{b} \ (\rm rad)$ & $2.18_{-0.06}^{+0.05}$ & $0.94_{-0.04}^{+0.04}$ & $1.56_{-0.28}^{+0.24}$ & -- & -- & $3.08_{-0.03}^{+0.01}$ & $1.58_{-0.40}^{+0.39}$ \\
$P_{c} \ (\rm days)$ & $11038.6_{-720.6}^{+1059.0}$ & -- & $5873.0_{-86.6}^{+93.4}$ & $769.4_{-2.9}^{+3.9}$ & $769.4_{-3.0}^{+3.8}$ & -- & -- \\
$K_{c} \ (\rm m\ s^{-1})$ & $105.8_{-2.1}^{+2.0}$ & -- & $17.2_{-0.6}^{+0.6}$ & $18.8_{-1.7}^{+1.9}$ & $18.6_{-1.7}^{+1.8}$ & -- & -- \\
$\sqrt{e_{c}} \sin \omega_{c}$ & $-0.37_{-0.06}^{+0.06}$ & -- & $-0.03_{-0.10}^{+0.12}$ & $-0.54_{-0.09}^{+0.09}$ & $-0.55_{-0.09}^{+0.10}$ & -- & -- \\
$\sqrt{e_{c}} \cos \omega_{c}$ & $-0.03_{-0.06}^{+0.06}$ & -- & $-0.21_{-0.05}^{+0.06}$ & $0.40_{-0.11}^{+0.10}$ & $0.38_{-0.11}^{+0.10}$ & -- & -- \\
$M_{0,c} \ (\rm rad)$ & $1.04_{-0.17}^{+0.20}$ & -- & $3.60_{-0.48}^{+0.49}$ & $5.68_{-0.28}^{+0.28}$ & $5.69_{-0.31}^{+0.27}$ & -- & -- \\
$\Omega_{c} \ (\rm rad)$ & $1.03_{-0.12}^{+0.10}$ & -- & $3.31_{-1.66}^{+1.69}$ & $0.73_{-0.40}^{+0.49}$ & $5.92_{-0.60}^{+0.23}$ & -- & -- \\
$i_{c} \ (\rm rad)$ & $1.98_{-0.08}^{+0.07}$ & -- & $1.63_{-0.34}^{+0.30}$ & $0.81_{-0.28}^{+0.49}$ & $2.59_{-0.32}^{+0.17}$ & -- & -- \\
$\sigma_{\rm CAR-VIS} \ (\rm m\ s^{-1})$ & -- & -- & $2.43_{-0.29}^{+0.34}$ & $9.59_{-0.69}^{+0.76}$ & $9.58_{-0.70}^{+0.80}$ & -- & -- \\
$\sigma_{\rm HARPS-N} \ (\rm m\ s^{-1})$ & -- & -- & $3.43_{-0.26}^{+0.30}$ & -- & -- & -- & -- \\
$\sigma_{\rm HARPS-pre} \ (\rm m\ s^{-1})$ & $3.08_{-0.24}^{+0.26}$ & $1.91_{-0.18}^{+0.21}$ & $1.88_{-0.23}^{+0.26}$ & -- & -- & -- & -- \\
$\sigma_{\rm HARPS-post} \ (\rm m\ s^{-1})$ & $3.90_{-1.24}^{+1.84}$ & $2.02_{-0.13}^{+0.14}$ & $4.25_{-0.51}^{+0.58}$ & -- & -- & $3.28_{-0.87}^{+1.01}$ & -- \\
$\sigma_{\rm HIRES-pre} \ (\rm m\ s^{-1})$ & -- & -- & $3.15_{-1.16}^{+1.29}$ & $21.51_{-5.66}^{+9.04}$ & $21.20_{-5.37}^{+8.66}$ & -- & -- \\
$\sigma_{\rm HIRES-post} \ (\rm m\ s^{-1})$ & -- & -- & $4.41_{-0.33}^{+0.37}$ & $9.33_{-1.05}^{+1.33}$ & $9.41_{-1.09}^{+1.32}$ & $7.11_{-2.02}^{+2.64}$ & -- \\
$\sigma_{\rm HRS} \ (\rm m\ s^{-1})$ & -- & -- & -- & -- & -- & -- & -- \\
$\sigma_{\rm PFS} \ (\rm m\ s^{-1})$ & $6.08_{-0.99}^{+1.28}$ & $3.04_{-0.31}^{+0.38}$ & $2.67_{-0.86}^{+1.46}$ & -- & -- & -- & $6.85_{-0.74}^{+0.89}$ \\
$\sigma_{\rm SOPHIE} \ (\rm m\ s^{-1})$ & -- & -- & -- & -- & -- & -- & -- \\
$\sigma_{\rm Tull} \ (\rm m\ s^{-1})$ & -- & -- & -- & -- & -- & -- & -- \\
$\sigma_{\rm UCLES} \ (\rm m\ s^{-1})$ & -- & $5.41_{-0.74}^{+0.86}$ & -- & -- & -- & -- & -- \\
$J_{\rm hip} \ (\rm mas)$ & $1.03_{-0.71}^{+1.06}$ & $0.78_{-0.54}^{+0.74}$ & $3.17_{-1.79}^{+1.66}$ & -- & -- & $22.17_{-2.62}^{+3.17}$ & $3.50_{-1.87}^{+1.53}$ \\
$S_{\rm gaia}$ & $1.07_{-0.05}^{+0.07}$ & $1.36_{-0.07}^{+0.07}$ & $1.20_{-0.08}^{+0.08}$ & $1.46_{-0.08}^{+0.07}$ & $1.45_{-0.07}^{+0.07}$ & $1.09_{-0.06}^{+0.08}$ & $1.07_{-0.05}^{+0.07}$ \\
$\Delta \alpha* \ (\rm mas)$ & $5.17_{-0.51}^{+0.39}$ & $0.95_{-0.10}^{+0.09}$ & $-0.30_{-0.83}^{+1.02}$ & $0.16_{-0.10}^{+0.09}$ & $-0.06_{-0.18}^{+0.14}$ & $-0.99_{-0.25}^{+0.29}$ & $1.33_{-1.50}^{+1.11}$ \\
$\Delta \delta \ (\rm mas)$ & $0.98_{-0.62}^{+0.74}$ & $0.80_{-0.09}^{+0.09}$ & $-0.57_{-0.54}^{+1.23}$ & $-0.28_{-0.15}^{+0.11}$ & $-0.31_{-0.11}^{+0.10}$ & $0.73_{-0.27}^{+0.23}$ & $2.18_{-1.13}^{+0.48}$ \\
$\Delta \varpi \ (\rm mas)$ & $-0.03_{-0.03}^{+0.03}$ & $-0.03_{-0.03}^{+0.03}$ & $-0.06_{-0.03}^{+0.03}$ & $-0.06_{-0.09}^{+0.09}$ & $-0.05_{-0.09}^{+0.09}$ & $-0.02_{-0.04}^{+0.04}$ & $-0.00_{-0.01}^{+0.01}$ \\
$\Delta \mu_{\alpha*} \ (\rm mas \ yr^{-1})$ & $0.07_{-0.07}^{+0.09}$ & $0.24_{-0.02}^{+0.03}$ & $0.11_{-0.08}^{+0.08}$ & $-0.18_{-0.10}^{+0.10}$ & $-0.02_{-0.10}^{+0.09}$ & $-0.06_{-0.07}^{+0.07}$ & $-0.09_{-0.06}^{+0.07}$ \\
$\Delta \mu_{\delta} \ (\rm mas \ yr^{-1})$ & $0.66_{-0.04}^{+0.04}$ & $-0.49_{-0.03}^{+0.03}$ & $0.12_{-0.06}^{+0.07}$ & $0.26_{-0.09}^{+0.08}$ & $0.33_{-0.08}^{+0.08}$ & $-0.54_{-0.15}^{+0.18}$ & $-0.13_{-0.04}^{+0.04}$ \\ \hline
\multicolumn{8}{c}{Derived parameters} \\ \hline
$i_{b} \ (\rm deg)$ & $125.1_{-3.3}^{+3.0}$ & $54.1_{-2.0}^{+2.1}$ & $89.7_{-16.0}^{+13.5}$ & -- & -- & $176.3_{-1.6}^{+0.8}$ & $90.7_{-23.1}^{+22.1}$ \\
$m_{b} \ (M_{\rm Jup})$ & $5.35_{-0.23}^{+0.24}$ & $0.94_{-0.04}^{+0.04}$ & $0.93_{-0.04}^{+0.06}$ & -- & -- & $31.32_{-9.65}^{+8.46}$ & $4.83_{-0.52}^{+0.63}$ \\
$q_{b}$ & $0.0082_{-0.0003}^{+0.0003}$ & $0.0020_{-0.0001}^{+0.0001}$ & $0.0019_{-0.0001}^{+0.0001}$ & -- & -- & $0.0628_{-0.0194}^{+0.0169}$ & $0.0074_{-0.0005}^{+0.0008}$ \\
$a_{b} \ (\rm AU)$ & $1.74_{-0.02}^{+0.02}$ & $3.58_{-0.06}^{+0.05}$ & $2.34_{-0.04}^{+0.03}$ & -- & -- & $0.36_{-0.01}^{+0.01}$ & $6.15_{-0.40}^{+0.46}$ \\
$e_{b}$ & $0.316_{-0.004}^{+0.004}$ & $0.013_{-0.009}^{+0.013}$ & $0.048_{-0.012}^{+0.012}$ & -- & -- & $0.332_{-0.010}^{+0.010}$ & $0.718_{-0.029}^{+0.032}$ \\
$\omega_{b} \ (\rm rad)$ & $1.49_{-0.02}^{+0.01}$ & $-0.87_{-1.80}^{+3.27}$ & $1.82_{-0.21}^{+0.21}$ & -- & -- & $-1.18_{-0.03}^{+0.03}$ & $-0.20_{-0.04}^{+0.04}$ \\
$T_{p,b} - 2450000 \ (\rm JD)$ & $7513.2_{-2.3}^{+2.3}$ & $7176.5_{-1402.3}^{+565.5}$ & $7819.5_{-66.5}^{+65.8}$ & -- & -- & $7839.6_{-0.4}^{+0.4}$ & $2828.8_{-692.3}^{+487.4}$ \\
$i_{c} \ (\rm deg)$ & $113.3_{-4.6}^{+3.7}$ & -- & $93.3_{-19.6}^{+17.3}$ & $46.5_{-16.2}^{+28.0}$ & $148.6_{-18.0}^{+9.9}$ & -- & -- \\
$m_{c} \ (M_{\rm Jup})$ & $9.25_{-0.36}^{+0.36}$ & -- & $0.95_{-0.05}^{+0.09}$ & $0.30_{-0.07}^{+0.13}$ & $0.40_{-0.13}^{+0.17}$ & -- & -- \\
$q_{c}$ & $0.0141_{-0.0005}^{+0.0005}$ & -- & $0.0020_{-0.0001}^{+0.0002}$ & $0.0019_{-0.0005}^{+0.0008}$ & $0.0026_{-0.0008}^{+0.0011}$ & -- & -- \\
$a_{c} \ (\rm AU)$ & $8.36_{-0.38}^{+0.53}$ & -- & $4.93_{-0.09}^{+0.09}$ & $0.88_{-0.04}^{+0.04}$ & $0.87_{-0.04}^{+0.04}$ & -- & -- \\
$e_{c}$ & $0.137_{-0.037}^{+0.047}$ & -- & $0.058_{-0.022}^{+0.024}$ & $0.467_{-0.076}^{+0.070}$ & $0.461_{-0.075}^{+0.067}$ & -- & -- \\
$\omega_{c} \ (\rm rad)$ & $-1.65_{-0.19}^{+0.15}$ & -- & $-2.39_{-0.54}^{+5.30}$ & $-0.94_{-0.20}^{+0.19}$ & $-0.96_{-0.19}^{+0.19}$ & -- & -- \\
$T_{p,c} - 2450000 \ (\rm JD)$ & $6095.1_{-251.8}^{+209.9}$ & -- & $4572.5_{-462.0}^{+443.6}$ & $7759.8_{-15.8}^{+15.6}$ & $7758.0_{-16.0}^{+15.6}$ & -- & -- \\
\end{tabular}}}
\end{table*}


\bibliography{planet}{}
\bibliographystyle{aasjournal}



\end{document}